\documentclass{elsart}
\usepackage{amstext}
\usepackage{graphicx}
\usepackage{epsfig}
\begin{document}
\begin{frontmatter}
\title{Shell model study of the isobaric chains $A=50$, $A=51$ and
  $A=52$}
\author{A. Poves and J. S\'anchez-Solano}
\address{Departamento de F\'{\i}sica Te\'orica C-XI, Universidad
  Aut\'onoma de Madrid, E--28049 Madrid, Spain}
\author{E. Caurier and F. Nowacki}
\address{Groupe de Physique Th\'eorique, Centre de Recherches
  Nucl\'eaires, Institut National de Physique Nucl\'eaire et de
  Physique des Particles, Centre National de la Recherche
  Scientifique, Universit\'e Louis Pasteur, Bo\^{\i}te Postale 28,
  F--67037 Strasbourg Cedex~2, France}
%\date{\today}
%\maketitle

\begin{abstract}
Shell model calculations in the full $pf$-shell are carried out for the
A=50, 51 and 52 isobars. The most frequently used effective
interactions for the $pf$-shell, KB3 and FPD6 are revisited and their
behaviour at the N=28 and Z=28 closures examined. Cures to their -relatively
minor- defaults are proposed, and a new mass dependent version
 called KB3G is released.
 Energy spectra, electromagnetic transitions and moments as well
as beta decay properties are computed and compared with
the experiment and with the results of the earlier interactions.
A high quality description is achieved. Other miscellaneous
topics are addressed; the Coulomb energy differences of the yrast
states of the mirror pair $^{51}$Mn-$^{51}$Fe and the systematics of the
magnetic moments of the N=28 isotones.
\end{abstract}
\end{frontmatter}

\section{Introduction}
\label{sec:intro}

The $pf$-shell has been the focus of a lot of activity in nuclear
structure during the last years. Prompted in some cases by the large
scale shell model results, that indicated the presence of a region of
deformation around $^{48}$Cr~\cite{a48,cr48,a4749}, a lot of new
experiments and  calculations have been
carried out, addressing many different issues; deformed bands
and  band
termination \cite{cr50e}, yrast traps \cite{fe52}, high K isomers
\cite{brando}, coexistence of deformed
bands of natural and non-natural parity \cite{poso}, effects of neutron proton
pairing \cite{poga}, etc. A very recent highlight has been the discovery of an
excited deformed band in $^{56}$Ni~\cite{rudolf} coexisting with the
spherical states based in the doubly magic ground state. In addition
to the exact shell model diagonalizations, the new Monte Carlo
techniques, SMMC and MCSM have been extensively applied to this
region~\cite{mc1,mc2,mc3,mc4}. Mean field
descriptions of various kinds have also been used to explore
different issues concerning this deformation region \cite{hara,terasaki}.

In this article we extend the full $pf$-shell calculations up to
A=52. Detailed results for $^{50}$Cr, $^{50}$Mn and $^{52}$Fe
using KB3 have been already published in refs \cite{fe52,cr50t,mn50} and
we will not deal with 
them here because the new interaction KB3G gives equivalent results.
 In the cases of $^{51}$Cr, $^{52}$Cr, $^{51}$Mn and
$^{52}$Mn we have carried out the full $pf$-shell calculation for the
yrast
states only. To perform detailed spectroscopy in the full space
would have demanded a huge amount of computer time, not justified by the
improvement on the results, as we have checked. Hence, for the
non-yrast states we shall present results in
a truncated (t=5) space (no more than 5 particles are allowed to
excite from the 1f$_{7/2}$ subshell). At this truncation level,
the most relevant states are sufficiently converged.

As we have discussed in detail elsewhere~\cite{a4749} our usual effective
interaction KB3~\cite{kb,pozu} produces a quasiparticle gap in
$^{56}$Ni about 1\,MeV too
large. Approaching the doubly magic closure, the effects of this
default become more visible. That  is the reason why, in a recent
study of the deformed excited band of $^{56}$Ni~\cite{rudolf}, we used
a  preliminary modified version of KB3
in order to have the correct gap. This modified version of KB3 was also 
used in the study of the M1 strength functions of the N=28 isotones in
ref.~\cite{m1n28}. We shall examine the interaction issues in
section II. In sections III, IV and V we present the spectroscopic
results for A=50, 51 and 52 respectively. In section VI we gather the
beta decay results. In section VII we discuss the Coulomb Energy
Differences (CED) between  the yrast states of the mirror pair
A=51. Finally, in section VIII we study the behaviour of the magnetic moments
of the N=28 isotones. We close with the
conclusions.

Throughout the paper $f$ stands for $f_{7/2}$ (except, of course, when
we speak of the $pf$ shell) and $r$, generically, for any or all of
the other subshells ($p_{1/2}\;p_{3/2}\;f_{5/2}$). Spaces of the type

\begin{equation}
f^{n-n_0} r^{n_0}+f^{n-n_0-1} r^{n_0+1}+\cdots+
f^{n-n_0-t} r^{n_0+t}
\end{equation}
represent possible truncations: $n_0$ is different from zero if more
than 8 neutrons (or protons) are present and when $t=n-n_0$ we have
the full space $(pf)^n$ for $A=40+n$.

The interaction KB3 is a (mostly) monopole modification of the original
Kuo-Brown one~\cite{kb}. The modifications are described in detail in \cite{a4749}.

In what follows, and unless specified otherwise, we use

\begin{itemize}
\item harmonic oscillator wave functions with $b=1.01 A^{1/6}$~fm;
\item bare electromagnetic factors in $M1$ transitions; effective
  charges of 1.5~$e$ for protons and 0.5~$e$ for neutrons in the
  electric quadrupole transitions and moments;
\item Gamow-Teller (GT) strength defined through
\begin{equation}
  B(GT)=\left(\frac{g_A}{g_V}\right)^2_{\mathrm{eff}} \langle
  \sigma\tau \rangle^2, \hspace{1cm}
  \langle \sigma\tau \rangle =\frac{\langle f||\sum_k
    \sigma^k t^k_\pm ||i\rangle}{\sqrt{2J_i+1}},
\end{equation}
where the matrix element is reduced with respect to the spin operator
only (Racah convention~\cite{edmons}), $\pm$ refers to $\beta^\pm$
decay, $t_\pm = (\tau_x \pm {\text{i}}\tau_y)/2$,
with $t_+ p = n$ and $(g_A/g_V)_{\mathrm{eff}}$ is the effective
axial to vector ratio for GT decays,
\begin{equation}
  \left(\frac{g_A}{g_V}\right)_{\mathrm{eff}} = 0.77
  \left(\frac{g_A}{g_V}\right)_{\mathrm{bare}},
\end{equation}
with $(g_A/g_V)_{\mathrm{bare}} = 1.2599(25)$~\cite{towner_last};

\item for Fermi decays we have
    \begin{equation}
     B(F)=\langle \tau \rangle^2, \hspace{1cm}
   \langle\tau\rangle =\frac{\langle f ||\sum_k
      t^k_\pm || i\rangle}{\sqrt{2J_i+1}};
    \end{equation}

\item half-lives, $T_{1/2}$, are found through

\begin{equation}
(f_A+f^\epsilon)\,T_{1/2}=\frac{6146\pm6}{(f_V/f_A)B(F)+B(GT)}.
\end{equation}
We follow ref.~\cite{wilki} in the calculation of the $f_A$ and $f_V$
integrals and ref.~\cite{bamby} for $f^\epsilon$. The experimental
energies are used.

\item The intrinsic quadrupole moments $Q_0$ are extracted from the
spectroscopic ones through

\begin{equation}
\label{eq:q0qsp}
Q_0=\frac{(J+1)\,(2J+3)}{3K^2-J(J+1)}\,Q_{\mathrm{spec}}(J), \hspace{0.5cm}
  \text{for $K\neq 1$}
\end{equation}
or from the BE2's through the rotational model prescription

\begin{equation}
\label{eq:q0be2}
B(E2,J\;\rightarrow\;J-2)=\frac{5}{16\pi}\,e^2|\langle JK20|
J-2,K\rangle|^2\, Q_0^2, \hspace{0.1cm} \text{for $K \neq \frac{1}{2},1$.}
\end{equation}

\end{itemize}

The diagonalizations are performed in the $m$-scheme using a fast
implementation of the Lanczos algorithm through the code {\sc
antoine}~\cite{antoine} or in  J-coupled scheme using the code
{\sc nathan} \cite{nathan}. Some details may be found in
ref.~\cite{masses}.  The strength functions are obtained using
Whitehead's prescription~\cite{white}, explained and illustrated in
refs.~\cite{cpz1,cpz2,bloom}. 

All the experimental results for which no explicit credit is given
come from the electronic version of Nuclear Data Sheets compiled by
Burrows \cite{nds}.

\section{The Interactions}
\label{sec:inter}

Following ref. \cite{duzu}, we shall treat the effective
interaction as a sum of a monopole and a multipole part. The monopole
part is responsible for the energies of the closed shells (CS) and the
closed shells plus or minus one particle (CS$\pm$1). In our valence
space, the starting point (or vacuum) is the $^{40}$Ca core, and the
single particle energies are provided by the levels of $^{41}$Ca. The
harmonic oscillator  closure should be $^{80}$Zr and its corresponding hole
states in  $^{79}$Y. Nevertheless, and taking care of the
effect of correlations, $^{48}$Ca and $^{56}$Ni can be also taken as
reference closed shells. In this we are lucky, because the information
given by $^{80}$Zr is rather useless. For N,Z$>$32 the influence of
the 1g$_{9/2}$ orbit is very strong and the $pf$ valence space is no
longer valid. Around $^{80}$Zr, the occupation of the orbits
1g$_{9/2}$ and 2d$_{5/2}$ drives the nuclei into deformed
shapes~\cite{rot,petro}. There are no experimental results available
for the single hole states in $^{79}$Y. Even if some excited levels were
accessible, a particle plus rotor spectrum should be expected, coming
from the coupling of the holes to the $^{80}$Zr deformed core.
With this information missing,  we have to rely on indirect
indications, as those coming from spherical Hartree Fock calculations
(not very accurate) or from the new monopole formulae fitted to the
real shell closures and their particle and hole partners~\cite{duflo}.
Hence, our main guidance to adjust the monopole part of our
interactions comes from the gaps around $^{48}$Ca and $^{56}$Ni and
from the single particle states  of $^{49}$Ca and $^{57}$Ni. The
multipole part of the interactions will be also analysed  in
terms of ``coherent'' multipoles, as proposed in
ref.~\cite{duzu}. This part of the interaction is 
left unchanged.  
 
The quasiparticle gap for $^{48}$Ca is defined as

\begin{equation}
  \label{gapn}
  \Delta = 2 BE(^{48}\text{Ca}) -BE(^{49}\text{Ca}) -
  BE(^{47}\text{Ca})
\end{equation}

\noindent
and similarly for $^{56}$Ni.

  In table~\ref{tab:gap1} we compare these quantities for the most popular
  effective interactions used in the $pf$-shell; KB3 \cite{kb,pozu}
  and FPD6 \cite{fpd6}. Both interactions use very similar
  single particle energies;
  $\epsilon_{7/2}$=0.0\,MeV, $\epsilon_{3/2}$=2.0\,MeV,
  $\epsilon_{1/2}$=4.0\,MeV and $\epsilon_{5/2}$=6.5\,MeV for KB3 and
  $\epsilon_{7/2}$=0.0\,MeV, $\epsilon_{3/2}$=1.89\,MeV,
  $\epsilon_{1/2}$=3.91\,MeV and $\epsilon_{5/2}$=6.49\,MeV for
  FPD6. The two body matrix elements defining FPD6 are scaled with the
  mass number A as  $(42/A)^{0.35}$, while KB3 does not incorporate any
  mass dependence.

\begin{table}[t]
\begin{center}
    \caption{$^{48}$Ca and  $^{56}$Ni gaps and single particle
      energies, with
      the interactions KB3 and FPD6 (energies in MeV).}
     \label{tab:gap1}
\vspace{0.3cm}
    \begin{tabular*}{\textwidth}{@{\extracolsep{\fill}}cccccccccc} 
\hline\hline
      &  \multicolumn{3}{c}{$\Delta$} & \multicolumn{3}{c}{$\epsilon_{\frac{5}{2}}-\epsilon_{\frac{3}{2}}$} & \multicolumn{3}{c}{$\epsilon_{\frac{1}{2}}-\epsilon_{\frac{3}{2}}$}\\
      \cline{2-4} \cline{5-7} \cline{8-10} 
    A=48  & Exp. & KB3 & FPD6 &  Exp. & KB3 & FPD6 
    & Exp. & KB3 & FPD6 \\

 \hline
      t=0 & 4.80 & 5.25 & 4.61 &  3.59 & 3.53 & 2.66 & 
    2.02 & 1.70 & 2.51 \\
      Full & 4.80 & 5.17 & 4.69 &  3.59 & 3.80 & 2.76 & 
    2.02 & 1.81 & 2.37 \\
 \hline
    A=56   & Exp. & KB3 & FPD6 & Exp. & KB3 & FPD6 &
     Exp. & KB3 & FPD6 \\
      t=0 & 6.39 & 8.57 & 7.41 &  0.77 & 0.38 & -0.48 &
    1.11 & 1.15 & 2.58 \\
      t=3 & 6.39 & 7.73 & 6.41 &  0.77 & 0.76 & 0.07 & 
    1.11 & 1.14 & 1.88 \\[2mm]
\hline\hline
    \end{tabular*}
\end{center}
\end{table}

   Notice that  KB3 does definitely better than FPD6 for the
   single particle spectra of $^{49}$Ca and $^{57}$Ni, while for the
    $^{56}$Ni gap, FPD6 is better.

   In addition the modification of KB3's gap, we want to 
    to make it mass dependent, in order to be able to use it safely
    around and beyond  $^{56}$Ni, therefore 
   we have to adjust the monopoles anew.
   As the gaps are subjet to variation when correlations are
   allowed, some trial and error fitting of the monopole changes is
   needed. The final modifications of KB3, defining KB3G for A=42 
   (we stick to the  $(42/A)^{1/3}$ mass dependence)
   are the following:
\begin{center}
\begin{eqnarray*}
  \label{eq:kb3g_def}
   V_{fp}^{T=1}\,(\text{KB3G}) = V_{fp}^{T=1}\,(\text{KB3})
   - 50\, \text{keV,}\\
   V_{fp}^{T=0}\,(\text{KB3G}) = V_{fp}^{T=0}\,(\text{KB3}) 
   - 100\, \text{keV,}\\
   V_{ff_{5/2}}^{T=1}\,(\text{KB3G}) =
   V_{ff_{5/2}}^{T=1}\,(\text{KB3}) 
   - 100\, \text{keV,}\\
   V_{ff_{5/2}}^{T=0}\,(\text{KB3G}) =
   V_{ff_{5/2}}^{T=0}\,(\text{KB3}) 
   - 150\, \text{keV,}\\
   V_{pp}^T(\text{KB3G}) = V_{pp}^T(\text{KB3}) 
   + 400\,\text{keV,}\\
\end{eqnarray*}
\end{center}
 where $p$ denotes any of the orbits $2p_{\frac{1}{2}}$ and
 $2p_{\frac{3}{2}}$.

The T=0 and T=1 modifications are different in order to recover
simultaneously the good gaps around $^{48}$Ca and $^{56}$Ni. The
modification of the $pp$ centroids bears no relationship with the
gaps, it is aimed to give a single hole spectrum of $^{80}$Zr in
accord with the predictions of ref \cite{duflo}. It is important to
remark that, for the nuclei already studied with KB3, KB3G 
produces equivalent results.

 In the case of FPD6, the gaps are nearly correct and the monopole
 defects amount to having the 1f$_{5/2}$ orbit too low and the 
 2p$_{1/2}$ orbit too high, both in  $^{49}$Ca and $^{57}$Ni. The
 modifications that repair that are:
\begin{center}
\begin{eqnarray*}
  \label{eq:fpd6*_def}
   V_{ff_{5/2}}^{T=0,1}\,(\text{FPD6*}) = V_{ff_{5/2}}^{T=0,1}\,
(\text{FPD6}) + 50\, \text{keV,}\\
   V_{fp_{1/2}}^{T=0,1}\,(\text{FPD6*}) = V_{fp_{1/2}}^{T=0,1}\,
(\text{FPD6}) - 50\, \text{keV}.
\end{eqnarray*}
\end{center}
  In table~\ref{tab:gap2} we present the resulting values around $^{48}$Ca and
  $^{56}$Ni. Notice that we have not aimed to an exact agreement. In
  particular, we know~\cite{fe54} that a t=3 calculations is not fully
 converged in $^{57}$Ni, therefore we have let room for the 1f$_{5/2}$
  to move up and for the 2p$_{1/2}$ to move  down a little. We also show the
  evolution of the quasiparticle neutron gaps through the N=28
  isotones for both modified interactions in table~\ref{tab:gap3},
  The gaps produced by FPD6 do not differ appreciably from those of FPD6*.
  The agreement with the experimental results is the best that can
  be reasonably aimed to.
\begin{table}
\begin{center}
    \caption{Same as in table~\ref{tab:gap1} with the modified
     interactions KB3G and FPD6*.}
     \label{tab:gap2}
\vspace{0.3cm}
    \begin{tabular*}{\textwidth}{@{\extracolsep{\fill}}cccccccccc} 
\hline\hline
      &  \multicolumn{3}{c}{$\Delta$} & \multicolumn{3}{c}{$\epsilon_{\frac{5}{2}}-\epsilon_{\frac{3}{2}}$} & \multicolumn{3}{c}{$\epsilon_{\frac{1}{2}}-\epsilon_{\frac{3}{2}}$}\\
      \cline{2-4} \cline{5-7} \cline{8-10}
      A=48 & Exp. & KB3G & FPD6* & Exp. & KB3G & FPD6* & Exp. & KB3G & FPD6* \\
 \hline
      t=0 & 4.80 & 4.73 & 4.61 & 3.59 & 3.20 & 3.04 & 2.02 & 1.71 & 2.13 \\
      Full & 4.80 & 4.69 & 4.68 & 3.59 & 3.44 & 3.10 & 2.02 & 1.82 & 2.03 \\
 \hline
      A=56 & Exp. & KB3G & FPD6* & Exp. & KB3G & FPD6* & Exp. & KB3G & FPD6* \\
      t=0 & 6.39 & 7.12 & 7.41 & 0.77 & 0.05 & 0.24 & 1.11 & 1.23 & 1.86 \\
      t=3 & 6.39 & 6.40 & 6.45 & 0.77 & 0.43 & 0.68 & 1.11 & 1.19 &
     1.45 \\[2mm] 
\hline\hline
    \end{tabular*}
\end{center}
\end{table}

\begin{table}
\begin{center}
 \caption{Theoretical quasiparticle neutron gaps of the N=28 isotones
   (in MeV) compared with experiment. For Z$>$23 we list t=3 results,
   for the rest, full $pf$-shell results (energies in MeV).}
 \label{tab:gap3}
\vspace{0.3cm}
 \begin{tabular*}{\textwidth}{@{\extracolsep{\fill}}cccccc}
\hline\hline
 & \multicolumn{3}{c}{$\Delta$ } &
 \multicolumn{2}{c}{$\delta=\Delta (exp) - \Delta (th)$}\\
\cline{2-4}\cline{5-6}
Nucleus & FPD6* & EXP. & KB3G & FPD6* & KB3G \\[0.2cm]
\hline
$^{48}$Ca & 4.68 & 4.81 & 4.70 & 0.13 & 0.11 \\
$^{49}$Sc & 4.05 & 4.07 & 3.99 & 0.02 & 0.08 \\
$^{50}$Ti & 4.66 & 4.57 & 4.41 & -0.09 & 0.16 \\
$^{51}$V  & 3.61 & 3.74 & 3.53 & 0.13 & 0.21 \\
$^{52}$Cr & 4.03 & 4.10 & 3.76 & 0.07 & 0.34 \\
$^{53}$Mn & 3.20 & 3.12 & 3.00 & -0.08 & 0.12 \\
$^{54}$Fe & 4.30 & 4.08 & 4.02 & -0.22 & 0.06 \\
$^{55}$Co & 4.00 & 4.01 & 4.16 & 0.01 & -0.15 \\
$^{56}$Ni & 6.45 & 6.39 & 6.41 & -0.06 & -0.02\\[2mm]
\hline\hline
\end{tabular*}
\end{center}
\end{table}

  Once the monopole part under control, we can analyse the multipole
  hamiltonian using the method of ref. \cite{duzu}. It amounts to
  diagonalize the monopole-free two body matrix elements in the
  particle-particle representation or in the particle-hole
  representation (i.e. after a Racah transformation). In the former
  case, the physically relevant terms are the isovector and isoscalar
  pairing terms, while in the latter all the multipoles will show up
  with different coherent strengths given by the lowest eigenvalues of
  the respective matrices. In table~\ref{tab:duzu}, we have listed
  these numbers 
  for the two interactions and for the most important multipoles.
  The differences are small; basically consist in
  FPD6 being 5-10$\%$ more intense than KB3 in all the
  channels. It is however satisfying the closeness of the values for
  two effective interactions derived by very different
  procedures. Remember that for its  multipole part,  KB3 is just
  equivalent to the original Kuo and Brown G-matrix \cite{kb}
  including the ``bubble'' correction, while FPD6 was obtained via a
  potential fit to selected energy levels in the $pf$-shell. This 
  supports of the conclusions in \cite{duzu} about the universality
  of the multipole shell model Hamiltonian.
\begin{table}
  \begin{center}
    \caption{Strengths (in MeV) of the coherent terms of the multipole
    Hamiltonian.}
    \label{tab:duzu}
\vspace{0.3cm}
    \begin{tabular*}{\textwidth}{@{\extracolsep{\fill}}cccccc} 
\hline\hline
  Interaction & \multicolumn{2}{c}{particle-particle} &
                \multicolumn{3}{c}{particle-hole} \\\cline{2-3}\cline{4-6}
                & JT=01  & JT=10 & $\lambda \tau$=20 &
                                 $\lambda \tau$=40 &
                                 $\lambda \tau$=11   \\ \hline
 KB3 & -4.75 & -4.46 & -2.79 & -1.39 & +2.46 \\
 FPD6 & -5.06 & -5.08 & -3.11 & -1.67 & +3.17 \\
 GOGNY & -4.07 & -5.74 & -3.23 & -1.77 & +2.46 \\[2mm]
\hline\hline
    \end{tabular*}
   \end{center}
 \end{table}

  We have purposely left to the end the line labeled GOGNY. It comes
  from the same analysis applied to the $pf$-shell two body matrix
  elements obtained using the density dependent interaction of Gogny
  \cite{gogny}. The calculation was carried out using the single
  particle wave function obtained in a spherical Hartree-Fock calculation
  for $^{48}$Cr in the uniform filling approximation in ref.~\cite{tomas}.
  In spite of the rather hybrid approach, the most important terms are
  again very similar to those arising from a G-matrix or from a shell
  model fit. In particular the agreement for the quadrupole and the
  spin-isospin terms is excellent. When it comes to pairing, this way of
  looking to the interaction can help to overcome some language barriers 
  between mean field  and shell model practitioners. As it is evident
  from the table, the Gogny interaction has essentially the same
  amount of isovector and isoscalar pairing than the realistic
  interactions. Therefore,
  it contains the right
   proton-neutron pairing. Thus, if there is something to
  blame in the mean field calculations for N=Z nuclei, it should be rather 
  the mean field approximations and not the interaction itself.

  In what follows we shall use mainly the interaction KB3G. We will
  compare in some cases with KB3 in order to evaluate the importance
  of the changes made in the monopole behaviour. The comparison with
  FPD6 will show to what extent the residual differences between good
  behaved interactions can have spectroscopic consequences. We have
  not attempted to compute all the states not even to draw those we have
  computed. Those not shown here can be obtained on request to
  the authors.

\section{The isobars A=50}
\label{sec:espectro50}

  We have carried out the full $pf$-shell calculations for all the
  isobars. The results for $^{50}$Cr and $^{50}$Mn have been already
  published in refs.~\cite{cr50t,cr50e,mn50}. The experimental data for which
  no specific credit is given are taken from ref.~\cite{nds}.

\subsection{Spectroscopy of $^{50}$Ca}
\label{sec:50ca}

\begin{figure}[h]
\begin{center}
    \epsfig{file=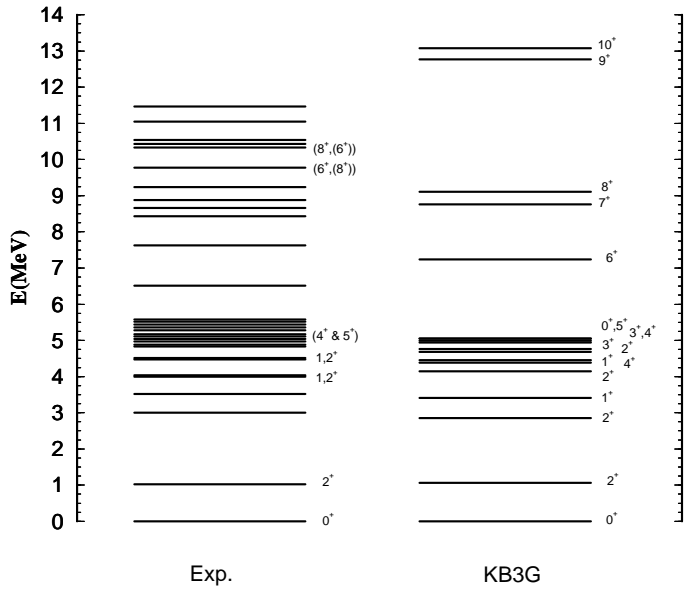,height=9cm}
    \caption{Energy levels of $^{50}$Ca.}
    \label{fig:fig1}
\end{center}
\end{figure}

 The experimental data are compared with the calculation in
 fig.~\ref{fig:fig1}. All the experimental states are plotted. We also
 plot all the calculated ones up to 5\,MeV. Beyond, only the  
 yrast states are drawn.  
 The scarcity of spin assignment
 precludes a more detailed analysis. The 2$^+$ excitation energy is
 well reproduced as well as the gap in the spectrum between 1 and 3\,MeV
 of excitation energy. KB3 gives equivalent results. No experimental 
 information on transitions is available. 

\begin{figure}[h]
\begin{center}
    \epsfig{file=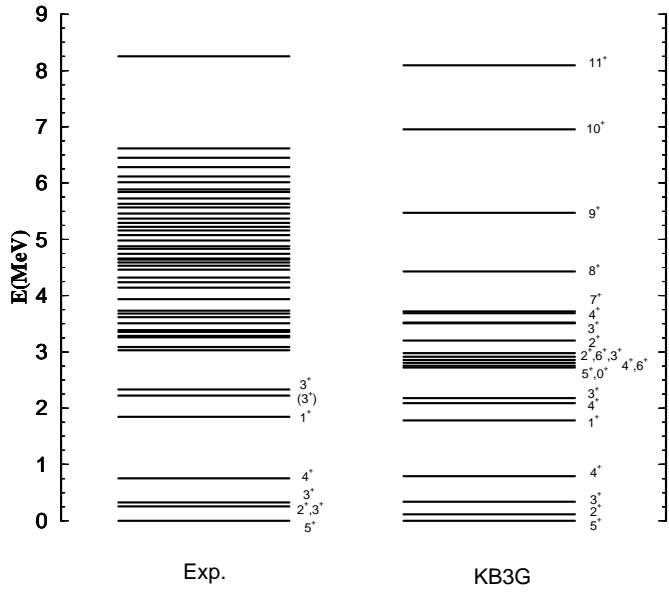,height=9cm}
    \caption{Energy levels $^{50}$Sc.}
    \label{fig:fig2}
\end{center}
\end{figure}

\subsection{Spectroscopy of $^{50}$Sc}
\label{sec:50sc}

 Figure~\ref{fig:fig2} reflects the experimental situation and our
 calculated spectrum. We have plotted all the experimental states even
 if only the lowest ones 
 have assigned spin. All the calculated states up to the first $7^+$ 
 state are shown. Above it, only the yrast band. The
 ground state multiplet, $(1f_{7/2})^9, 2p_{3/2}$ is well reproduced.
 The first excited state must
 be a 2$^+$ because the multiplet cannot contain two 3$^+$'s. The
 triplet at $\sim$2~MeV, belonging to the configurations
 $(1f_{7/2})^9, (1f_{5/2}, 2p_{1/2})^1$,
 is also well given. Beyond, the level density
 increases rapidly and no spin assignments are available.
 Notice that  the experimental density of states is nicely reproduced 
 up to 4~MeV.

\subsection{Spectroscopy of $^{50}$Ti}
\label{sec:50ti}

\begin{figure}[t]
\begin{center}
    \epsfig{file=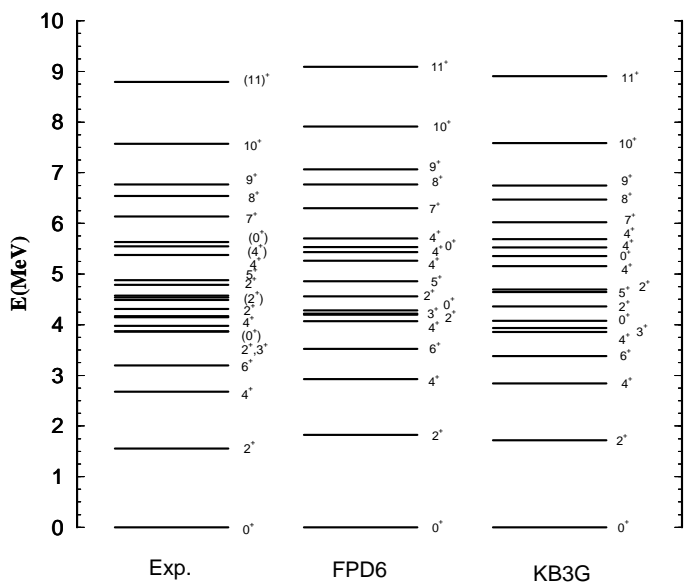,height=9cm}
    \caption{Energy levels of $^{50}$Ti.}
    \label{fig:fig3}
\end{center}
\end{figure}

$^{50}$Ti is the stable member of the isobaric multiplet and the
best known experimentally. It is semi-magic and therefore any
defects in the neutron gap would be evident in the spectrum.  In
fig~\ref{fig:fig3} the experimental level scheme is compared with the
full pf-shell results using the interactions KB3G and FPD6.  Up to 4\,
MeV all the states are plotted. Between 4\,MeV and 5.6\,MeV only those with
unambiguous experimental spin assignment and beyond, only the yrast.
The agreement is impressive for the two interactions, perhaps with a
bonus for KB3G. Notice the excellent reproduction of the second and
third 0$^+$ states as well as the perfect location of the high spin
states. The second 0$^+$  corresponds to the excitation of two
neutrons to  the 2$p_{\frac{3}{2}}$ orbit, while the third  has a
more complex structure. 

\begin{table}[h]
\begin{center}
    \caption{Transitions in $^{50}$Ti.}
    \label{tab:ti50}
\vspace{0.3cm}
\begin{tabular*}{\textwidth}{@{\extracolsep{\fill}}ccc}
\hline\hline
$B(E2)$ ($e^2\,fm^4$) & Exp. & Th. \\
\hline
$2^+\rightarrow 0^+$ & 58(9) & 88 \\
$4^+\rightarrow 2^+$ & 60(1) & 86 \\
$6^+\rightarrow 4^+$ & 34(1) & 41 \\[2mm]
\hline\hline
\end{tabular*}
\end{center}
\end{table}

The quadrupole and the magnetic moments of the yrast $J=2^+$ 
and $J=6^+$ states are known~\cite{toi}.
Their values,  $\mu_{exp}(2^+)=2.89(15)\mu_N$~\cite{speidel},
$Q_{exp}(2^+)=+8(16)\,e\,fm^2$ and  
$\mu_{exp}(6^+)=+9.3(10)\mu_N$ agree quite well with our predictions;
$\mu(2^+)=+2.5\mu_N$, $Q(2^+)=+6\,e\,fm^2$ and
$\mu(6^+)=+8.3\mu_N$. A more detailed discussion on the magnetic
moments is given in section~\ref{sec:mu}. Similarly the 
$E2$ transitions of the yrast $J=2^+\text{,}\,4^+$ y $6^+$ states are well
reproduced by the calculation (see table~\ref{tab:ti50}).

\subsection{Energy levels of $^{50}$V}
\label{sec:50v}

\begin{figure}[h]
\begin{center}
    \epsfig{file=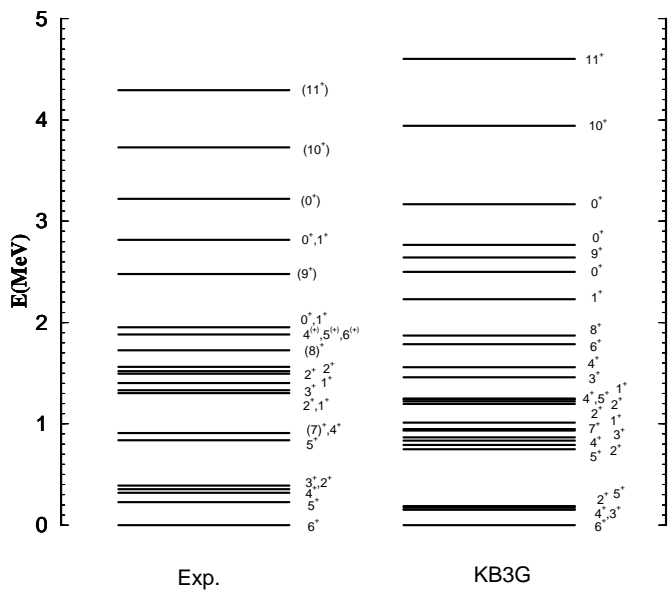,height=9cm}
    \caption{Energy levels of $^{50}$V.}
    \label{fig:fig4}
\end{center}
\end{figure}

 The experimental information is also very rich for this
 nucleus. Given the high level density, we have represented in
 fig.~\ref{fig:fig4} all the states  up to 1.6 MeV only. Above,
 just the high spins and a couple of 0$^+$'s and 1$^+$'s that may have 
 experimental counterparts.
 The calculation gives a good reproduction of the ground
 state quintuplet and locates correctly the high spin states
 9$^+$, 10$^+$ and 11$^+$. The only discrepancy affects to the bunch of
 1$^+$, 2$^+$ and 3$^+$ states at around 1.5 MeV that are placed
 500\,keV too low.

 The magnetic and the quadrupole moment of
the $J=6^+$ ground state have been measured~\cite{toi}. Their values
($\mu_{exp}(6^+)=+3.3456889(14)\mu_N$,
$Q_{exp}(6^+)=+20.9(40)\,e\,fm^2$) are very well reproduced by the
calculations 
($\mu(6^+)=+3.18\mu_N$, $Q(6^+)=+19.6\,e\,fm^2$). The electromagnetic
transitions of the yrast states are also known (table~\ref{tab:v50}).
 Notice the excellent agreement found for the -dominant- M1
 transitions. For the $\Delta$J=1, E2 transitions the huge error bars
 make the comparison meaningless. On the contrary, for the only
 $\Delta$J=2 transition measured, the agreement is very good.

\begin{table}[h]
\begin{center}
    \caption{Transitions in $^{50}$V.}
    \label{tab:v50}
\vspace{0.3cm}
\begin{tabular*}{\textwidth}{@{\extracolsep{\fill}}ccc}
\hline\hline
& Exp. & Th.\\
\hline
$B(M1)$ & ($\mu_N^2$) & ($\mu_N^2$) \\
$7^+\rightarrow 6^+$ & 1.2(2) & 1.0 \\
$8^+\rightarrow 7^+$ & 0.3(1) & 0.2 \\
$11^+\rightarrow 10^+$ & 0.9(3) & 1.1 \\[0.8mm]
\hline
$B(E2)$ & ($e^2\,fm^4$) & ($e^2\,fm^4$) \\
$7^+\rightarrow 6^+$ & $875^{+1313}_{-875}$ & 108 \\
$8^+\rightarrow 6^+$ & 98(44) & 74 \\
$8^+\rightarrow 7^+$ & $219^{+438}_{-219}$ & 10 \\
$11^+\rightarrow 10^+$ & $109^{+328}_{-109}$ & 29\\[2mm]
\hline\hline
\end{tabular*}
\end{center}
\end{table}

\section{The isobars A=51}
\label{sec:espectro51}

\subsection{Spectroscopy of $^{51}$Ca}
\label{sec:51ca}

\begin{figure}[h]
\begin{center}
    \epsfig{file=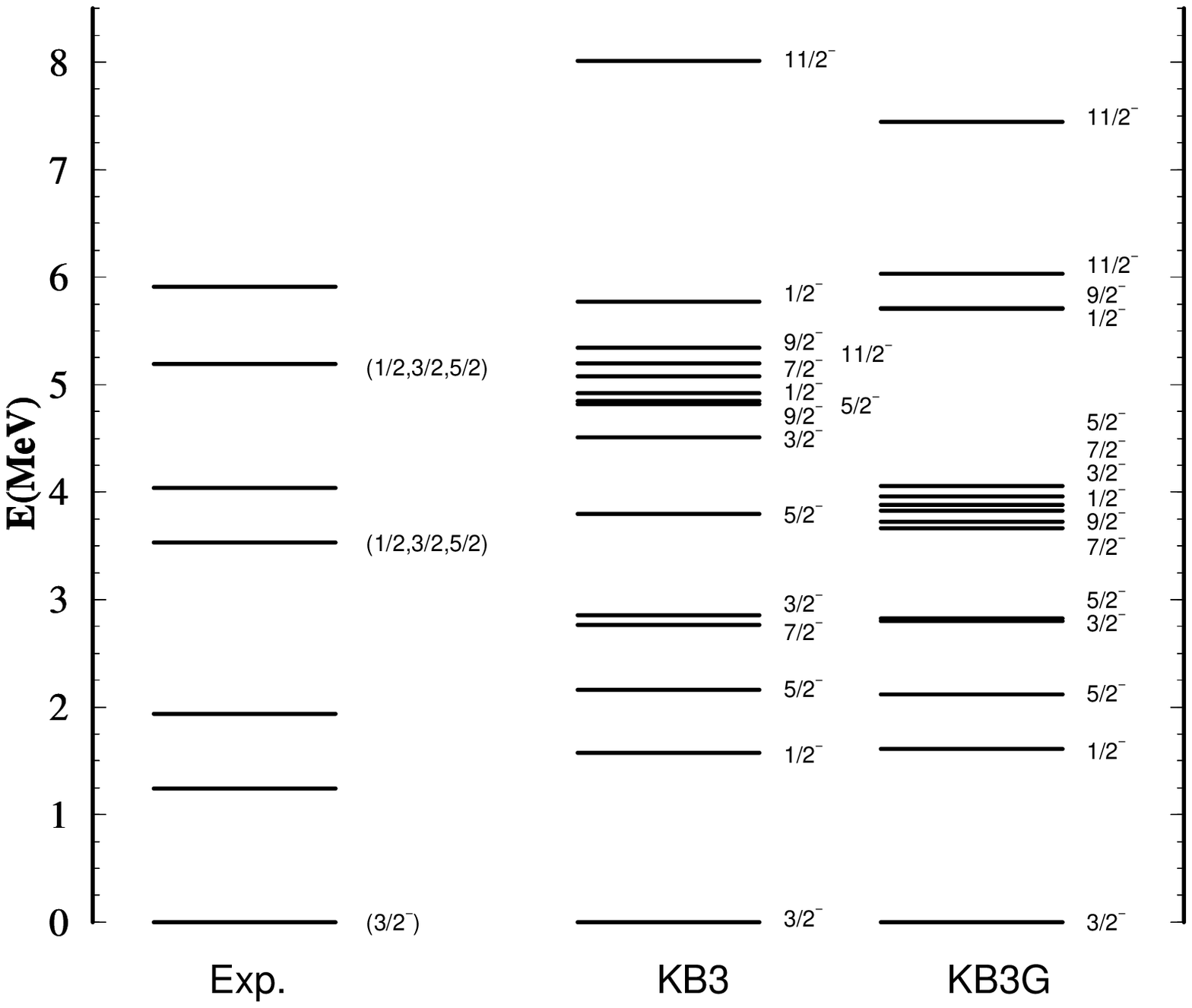,height=9cm}
    \caption{Energy levels of $^{51}$Ca.}
    \label{fig:fig5}
\end{center}
\end{figure}

The experimental data become rarer as we go far from
stability. In $^{51}$Ca only the ground state $J=\frac{3}{2}^-$ is
assigned in the experimental scheme. It is what is expected in any
reasonable calculation. This nucleus was also studied in
ref.~\cite{novo1}, using the interactions KB3 and FPD6. As
no comparison with the data is possible, we can examine the
predictions of the different interactions.  In fig.~\ref{fig:fig5} we
compare KB3 and KB3G. As the gap is very similar for the two
interactions in this nucleus, the differences must be due to the
changes in the $pp$ interaction. This is clearly seen in the figure.
The four states 3/2$^-$, 1/2$^-$, 5/2$^-$ and 3/2$^-$ are dominantly
$p^3$ states and remain unchanged between KB3 and KB3G. However, the
states 7/2$^-$ and 5/2$^-$ are swapped with a relative change of
about 2\,MeV. This is due to their different structures; the 5/2$^-$ is
$f_7^8 p^2 f_5$, thus, with KB3G it gains 800\,keV relative
 to the $f_7^8 p^3$ states,
the 7/2$^-$ has a structure $f_7^7 p^4$ and it looses about 1.2\,MeV.
Similar arguments explain the lowering of the bunch of states at about
5\,MeV excitation energy. Therefore a better experimental spectrum will
be of much help in refining these monopolar changes.

\subsection{Spectroscopy of $^{51}$Sc}
\label{sec:51sc}

\begin{figure}[h]
\begin{center}
    \epsfig{file=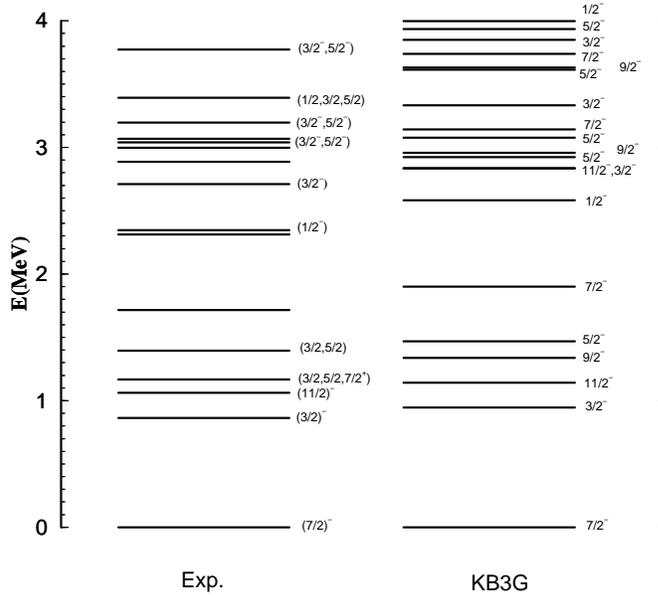,height=9cm}
    \caption{Energy levels of $^{51}$Sc.}
    \label{fig:fig6}
\end{center}
\end{figure}

As in $^{51}$Ca, this nucleus was studied in ref.~\cite{novo1}. Some
numerical errors present in this reference were discussed and corrected in 
refs.~\cite{novo2,novo3,novo4}. Figure~\ref{fig:fig6} shows all the
known experimental states. The calculated ones up to 4\,MeV energy are
also shown. Our results using KB3G are slightly better
than those obtained with KB3 and substantially better than those produced by 
FPD6~\cite{novo2,novo4}, specially up to 2\,MeV of excitation energy
where the spacing of the calculated levels follows closely that of the 
experiment.

\subsection{Spectroscopy of $^{51}$Ti}
\label{sec:51ti}

\begin{figure}[h]
\begin{center}
    \epsfig{file=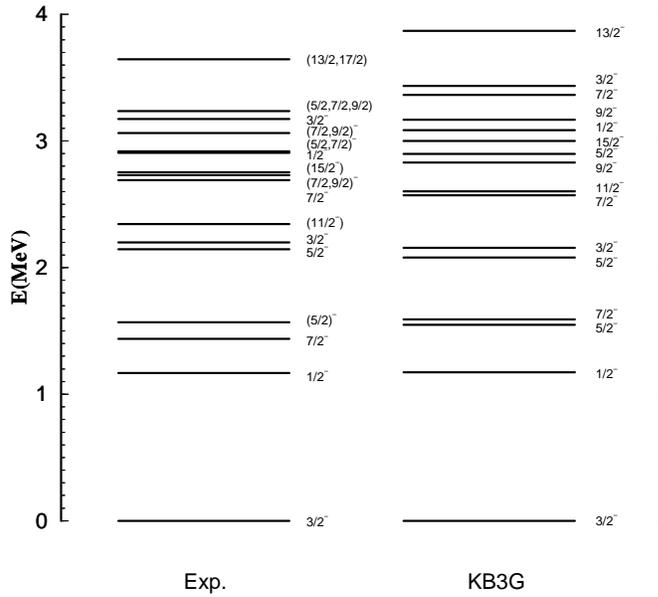,height=9cm}
    \caption{Energy levels of $^{51}$Ti.}
    \label{fig:fig7}
\end{center}
\end{figure}

  In figure~\ref{fig:fig7} we have plotted all the levels up to 3\,MeV
  and a few more up to 4\,MeV 
  which have spin and parity assignments. No high spin states are
  known beyond this energy. The agreement is quite good, exception
  made of a couple of doublets that come out inverted ($J=\frac{5}{2}^-$ and
  $J=\frac{7}{2}^-$ at $1.5$\,MeV and the second $J=\frac{7}{2}^-$
  with the first $J=\frac{11}{2}^-$ at around $2.5$\,MeV). The
  experimental spin assignment for the state around 3.7\,MeV
  is either $J=\frac{13}{2}^-$ or $J=\frac{17}{2}^-$. Our calculation
  predicts a  $J=\frac{13}{2}^-$ close to this energy, while the first
  $J=\frac{17}{2}^-$
  state is predicted 2\,MeV higher. If we
  compute the  $^{51}$Ti spectrum with FPD6 there are not significant
  differences, because the dominant configurations are 
  $(f_{\frac{7}{2}}^8\,p_{\frac{3}{2}}^1)_{\nu}\,(f_{\frac{7}{2}}^2)_{\pi}$,
  therefore the $f_{\frac{5}{2}}$ does not play an important role. We
  can interpret most of the  $^{51}$Ti level scheme 
  as the result of the coupling 
  $^{50}$Ti$\otimes p_{\frac{3}{2}}^1$. 

 Some electromagnetic transitions are experimentaly known that are well
 reproduced by the calculation (see table~\ref{tab:ti51}). Again in this case,
 the E2 transitions with  $\Delta$J=1 are poorly determined
 experimentally, due to the
 dominance of the M1 transitions. The calculated  $\Delta$J=2 B(E2)'s
 agree extremely well with the measured values.
\begin{table}[h]
\begin{center}
    \caption{Transitions in $^{51}$Ti.}
    \label{tab:ti51}
\vspace{0.3cm}
\begin{tabular*}{\textwidth}{@{\extracolsep{\fill}}ccc}
\hline\hline
& Exp. & Th.\\
\hline
$B(M1)$ & ($\mu_N^2$) & ($\mu_N^2$) \\
$\frac{5}{2}^-\rightarrow \frac{3}{2}^-$ & 0.06(2) & 0.07 \\[0.8mm]
\hline
$B(E2)$ & ($e^2\,fm^4$) & ($e^2\,fm^4$) \\
$\frac{5}{2}^-\rightarrow \frac{3}{2}^-$ & 348(112) & 88 \\
$\frac{7}{2}^-\rightarrow \frac{3}{2}^-$ & 225(202) & 88 \\
$\frac{11}{2}^-\rightarrow \frac{7}{2}^-$ & 95(17) & 95 \\
$\frac{15}{2}^-\rightarrow \frac{11}{2}^-$ & 62(24) & 53\\[2mm]
\hline\hline
\end{tabular*}
\end{center}
\end{table}

\subsection{Spectroscopy of $^{51}$V}
\label{sec:51v}

 $^{51}$V is the stable isotope in the isobar chain $A=51$ and
 corresponds to the N=28 neutron shell closure. The experimental
 information is very rich, extending up to
 $13$\,MeV. Figure~\ref{fig:fig8} shows the experimental yrast band
 with all the known high spin assignments compared with the
 calculated one including three more predicted spins. The agreement is
 extremely good. For a more complete analysis below 3\,MeV energy, where
 more experimental information is available, we have plotted
 in figure~\ref{fig:fig9} all the levels up to that energy and a few more
 above that can be put in correspondence with the experiment.
 We have calculated with KB3G and FPD6, both
 reproduce nicely the level scheme but definitely the quality of the
 agreement with KB3G is better.
   The magnetic moments of the ground state  $J=\frac{7}{2}^-$ and
  first excited state $J=\frac{5}{2}^-$ as well as the quadrupole
  moment of the ground state are known~\cite{toi}. Their values are:

\begin{figure}[h]
\begin{center}
    \epsfig{file=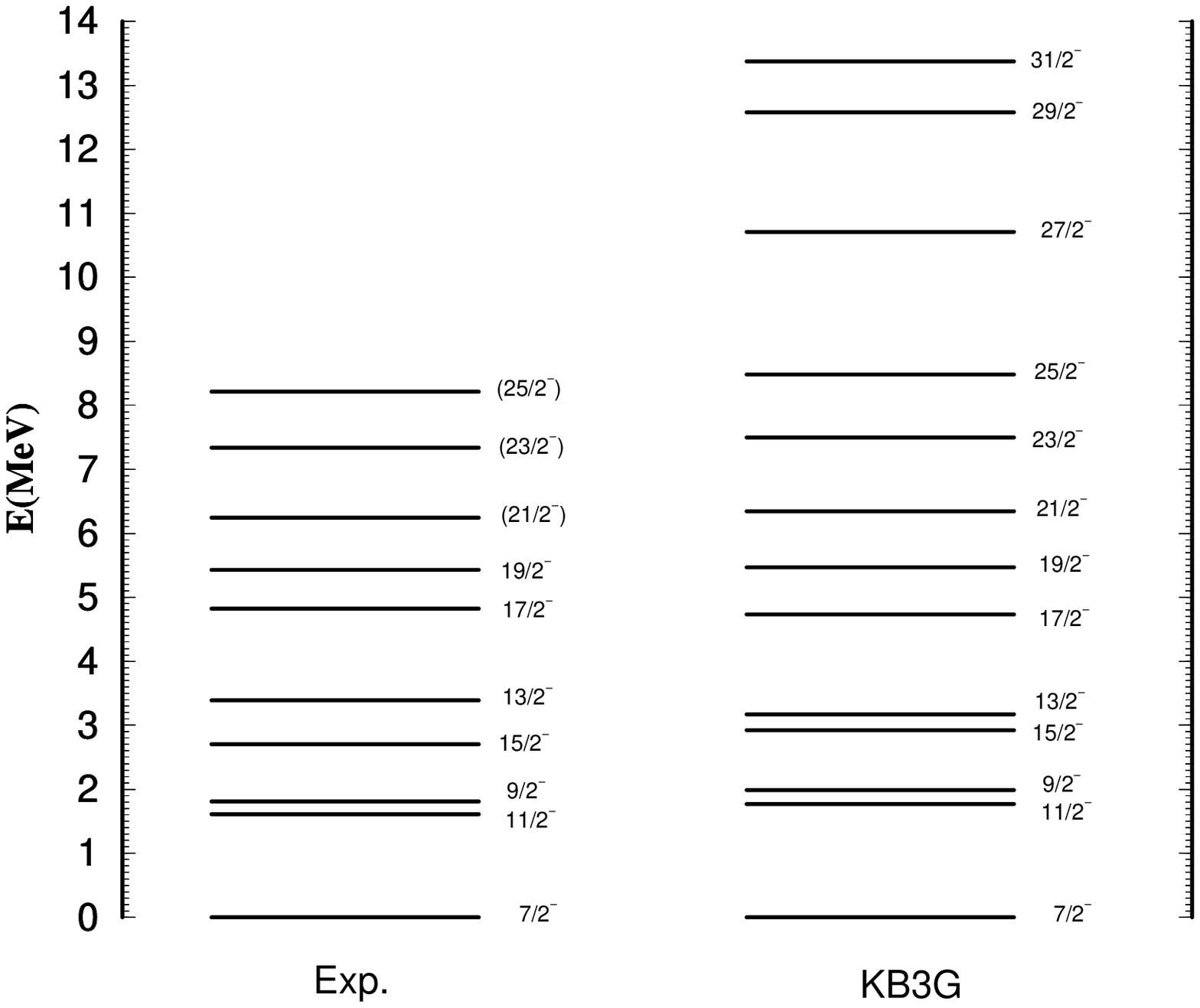,height=9cm}
    \caption{Yrast band of $^{51}$V.}
    \label{fig:fig8}
\end{center}
\end{figure}

\begin{figure}[t]
\begin{center}
    \epsfig{file=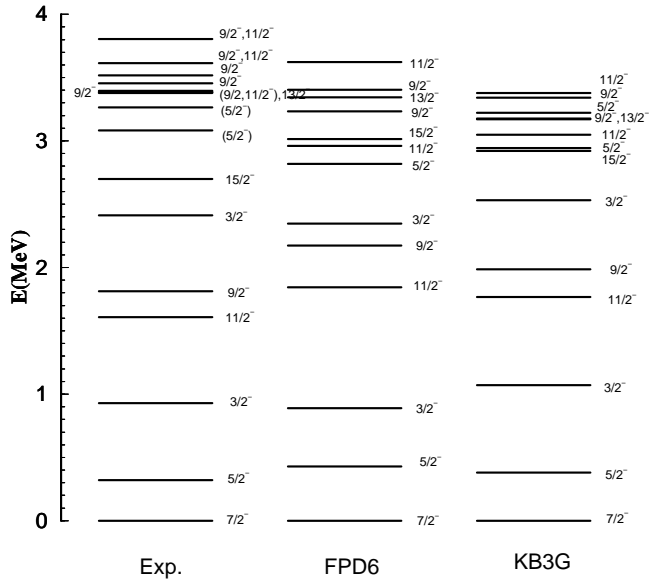,height=9cm}
    \caption{Energy levels of $^{51}$V.}
    \label{fig:fig9}
\end{center}
\end{figure}

\begin{center}
$\mu_{exp}(\frac{7}{2}^-)=+5.14870573(18)\,\mu_N$ \,\,\,\,
$Q_{exp}(\frac{7}{2}^-)=-5.2(10)\,e\,fm^2$\\
$\mu_{exp}(\frac{5}{2}^-)=+3.86(33)\,\mu_N$
\end{center}

\noindent
while the calculation gives:

\begin{center}
$\mu(\frac{7}{2}^-)=+4.99\,\mu_N$  \,\,\,\,
$Q(\frac{7}{2}^-)=-6.5\,e\,fm^2$\\
$\mu(\frac{5}{2}^-)=+3.36\,\mu_N$\\
\end{center}

\noindent
 in good agreement with the data. Furthermore, 
 some electromagnetic transitions,
 mainly between yrast states are known experimentally ~\cite{toi}.
 The calculation does also quite well, as it is shown in table~\ref{tab:v51}.
\begin{table}[h]
\begin{center}
    \caption{Transitions in $^{51}$V.}
    \label{tab:v51}
\vspace{0.3cm}
\begin{tabular*}{\textwidth}{@{\extracolsep{\fill}}ccc}
\hline\hline
& Exp. & Th.\\
\hline
$B(M1)$ & ($\mu_N^2$) & ($\mu_N^2$) \\
$\frac{9}{2}^-\rightarrow \frac{7}{2}^-$ & 0.0006(2) & 0.2$\cdot 10^{-6}$ \\
$\frac{13}{2}^-\rightarrow \frac{11}{2}^-$ & $<$0.0077 & 0.014 \\[0.8mm]
\hline
$B(E2)$ & ($e^2\,fm^4$) & ($e^2\,fm^4$) \\
$\frac{9}{2}^-\rightarrow \frac{7}{2}^-$ & 35(6) & 32 \\
$\frac{11}{2}^-\rightarrow \frac{7}{2}^-$ & 95(8) & 103 \\
$\frac{13}{2}^-\rightarrow \frac{11}{2}^-$ & $<$34.8 & 0.01 \\
$\frac{15}{2}^-\rightarrow \frac{11}{2}^-$ & 66(6) & 78\\[2mm]
\hline\hline
\end{tabular*}
\end{center}
\end{table}

\subsection{Spectroscopy of $^{51}$Cr}
\label{sec:51cr}

\begin{figure}[b]
\begin{center}
    \epsfig{file=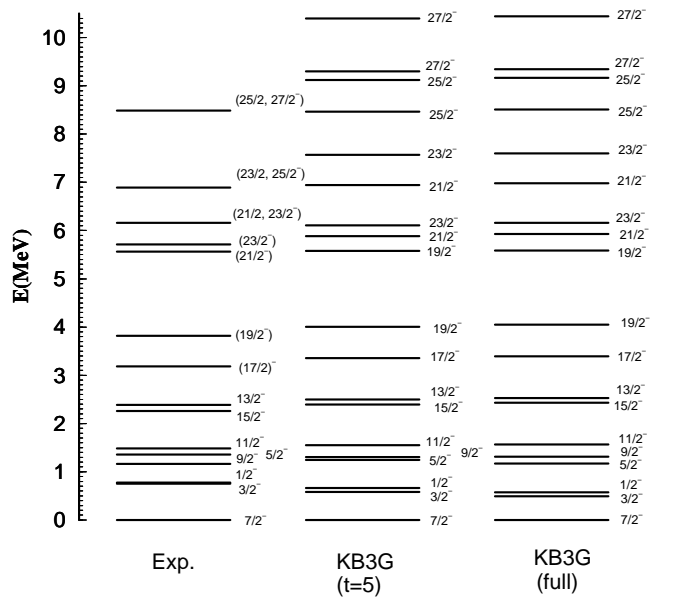,height=9cm}
    \caption{Yrast band of $^{51}$Cr; experiment,  $t=5$
     and full calculation.}
    \label{fig:fig10}
\end{center}
\end{figure}

\begin{figure}[h]
\begin{center}
    \epsfig{file=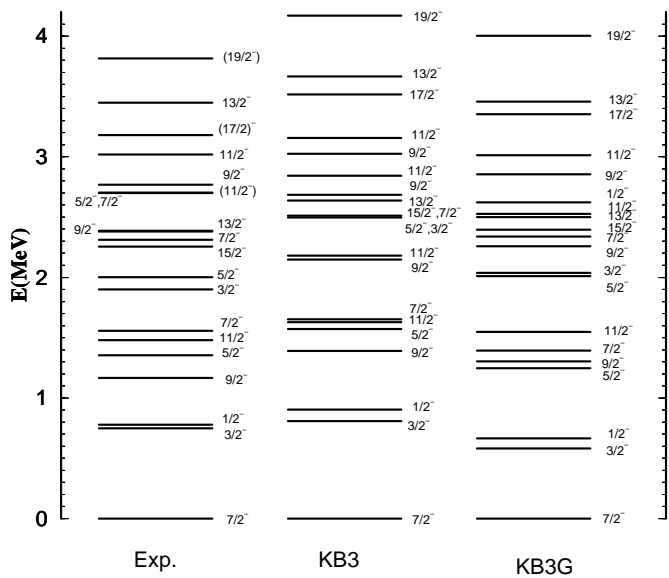,height=9cm}
    \caption{Energy levels of $^{51}$Cr, $t=5$.}
    \label{fig:fig11}
\end{center}
\end{figure}

 We present in fig.~\ref{fig:fig10} the results for the yrast band
 of $^{51}$Cr in a  $t=5$ truncation and in the full space using
 KB3G. We have also included the low-lying $J=\frac{1}{2}^-,\frac{3}{2}^-$
 and  $\frac{5}{2}^-$ states, and a second
  state of each spin from $J=\frac{19}{2}^-$ on. The figure
 shows  that the full calculation do not
 modify  substantially  the results obtained at  $t=5$.
 The agreement with the
 experimental data is very satisfactory. Notice that our prediction for the
 spins of the triplet of states around 6~MeV is shifted down by one unit
 relative to the preliminary experimental assignment, that should be
 revised in the light of our results.
 The detailed comparison with all the
 levels  up to 2.8\,MeV, and with the high
 spin states above this energy up to $J=\frac{19}{2}^-$ is carried out
 with a $t=5$ calculation. The results for both KB3 and KB3G are
 plotted in figure~\ref{fig:fig11}.

  The agreement is very good in spite of some local inversions. We can
  also notice that KB3G improves  systematically  the KB3 results.
  The sequence of states 
   $J=\frac{7}{2}^-,\frac{9}{2}^-,J=\frac{11}{2}^-$
 belong to the configuration
  $(1f_{\frac{7}{2}})^{11}$, while the doublets
 $J=\frac{3}{2}^-,\frac{1}{2}^-$ and  $J=\frac{5}{2}^-,\frac{7}{2}^-$
 located at 800\,keV and  1.5\,MeV, correspond to a neutron jump to the
  orbit $2p_{\frac{3}{2}}$, a sort of coupling 
  $^{50}$Cr$\otimes p_{\frac{3}{2}}^1$. As a consequence of the
  reduction of the gap, KB3G put these doublets at lower energy, while
  leaving the yrast ones unchanged. This provides a good illustration
  of a direct monopole effect. A similar though more pronounced effect
  is seen in the lowering of the doublet
  $J=\frac{3}{2}^-,\frac{5}{2}^-$ experimentally at 2\,MeV predicted
  by  KB3 \, 500\,keV too high, again a gap effect solved by
  KB3G. Besides, the high spin states were also too high in KB3 and come
  now very close to their experimental position.
  
  The magnetic moments of the ground state and the first excited state
  are known,
   ($\mu_{exp}(\frac{7}{2}^-)=-0.934(5)\,\mu_N$ and
 $\mu_{exp}(\frac{3}{2}^-)=-0.86(12)\,\mu_N$). Our prediction for the
  ground state is quite accurate 
 $\mu(\frac{7}{2}^-)=-0.886\,\mu_N$ while for the 
 $J=\frac{3}{2}^-$ state our result $\mu(\frac{3}{2}^-)=-0.40\,\mu_N$
  is off by a factor two.

   In table~\ref{tab:cr51} we compare the experimental and calculated
   transition probabilities. The B(M1)'s and 
   the $\Delta$J=2 B(E2)'s  are well reproduced .
    However this is not the case for the 
   $\Delta$J=1 B(E2)'s, in particular the discrepancies are huge for
   the transitions  $\frac{11}{2}^- \rightarrow  \frac{9}{2}^-$  and
   $\frac{13}{2}^- \rightarrow  \frac{11}{2}^-$. We have carried out a
   consistency test using the experimental M1/E2 branchings, $\delta$,
   and we have found that the  experimental  $\delta$ values are
   inconsistent with the experimental B(E2)'s but consistent with the
   calculated values. This is not surprising, because of the
   complete dominance of the M1 transition, that may cause large errors
   in the extraction of the B(E2) value. In order to discard any other
   origin of the discrepancy we have repeated the calculation with
   FPD6 and the situation is the same or even worse.
\begin{table}[h]
\begin{center}
    \caption{Transitions in $^{51}$Cr.}
    \label{tab:cr51}
\vspace{0.3cm}
\begin{tabular*}{\textwidth}{@{\extracolsep{\fill}}ccc}
\hline\hline
& Exp. & Th.\\
\hline
$B(M1)$ & ($\mu_N^2$) & ($\mu_N^2$) \\
$\frac{9}{2}^-\rightarrow \frac{7}{2}^-$ & 0.31683(3043) & 0.439 \\
$\frac{11}{2}^-\rightarrow \frac{9}{2}^-$ & 1.253(537) & 1.295 \\
$\frac{13}{2}^-\rightarrow \frac{11}{2}^-$ & 0.8950(1969) & 1.356 \\[0.8mm]
\hline
$B(E2)$ & ($e^2\,fm^4$) & ($e^2\,fm^4$) \\
$\frac{9}{2}^-\rightarrow \frac{7}{2}^-$ & 124(56) & 213 \\
$\frac{11}{2}^-\rightarrow \frac{7}{2}^-$ & 67(34) & 72 \\
$\frac{11}{2}^-\rightarrow \frac{9}{2}^-$ & 8(3) & 180 \\
$\frac{13}{2}^-\rightarrow \frac{11}{2}^-$ & 6(1) & 151 \\
$\frac{15}{2}^-\rightarrow \frac{11}{2}^-$ & 44(1) & 53\\[2mm]
\hline\hline
\end{tabular*}
\end{center}
\end{table}

\subsection{Spectroscopy of $^{51}$Mn}
\label{sec:51mn}

\begin{figure}[h]
\begin{center}
    \epsfig{file=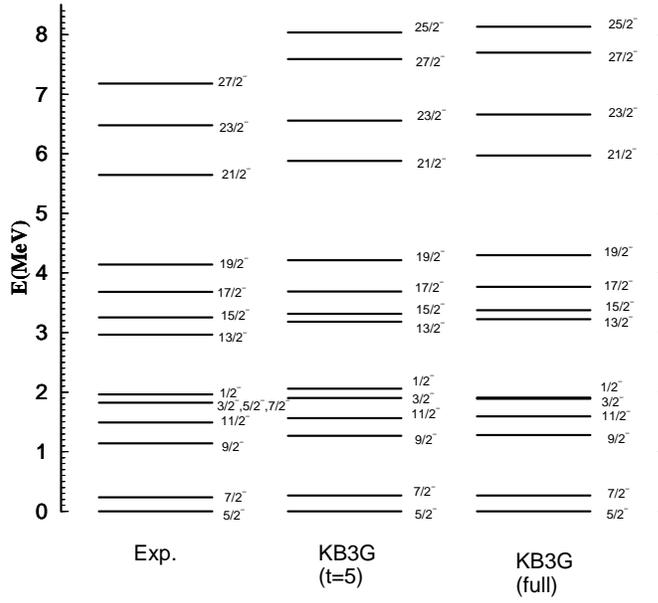,height=9cm}
    \caption{Yrast band of  $^{51}$Mn; experiment,  $t=5$
     and  full calculation.}
    \label{fig:fig12}
\end{center}
\end{figure}

\begin{figure}[h]
\begin{center}
    \epsfig{file=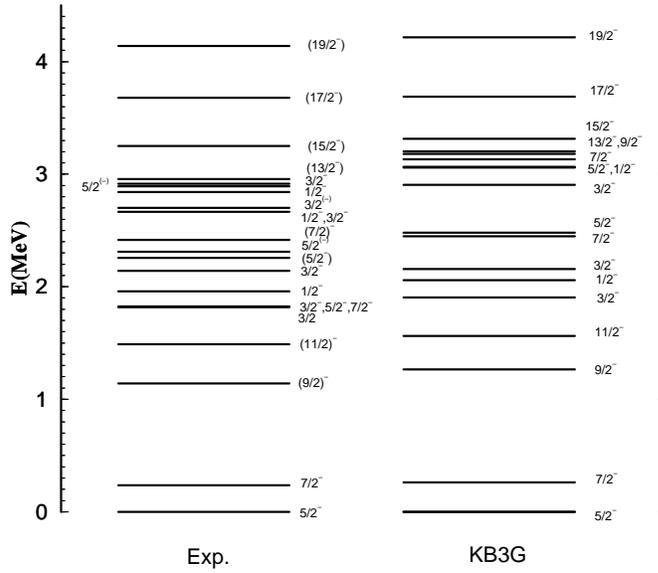,height=9cm}
    \caption{Energy levels of $^{51}$Mn, $t=5$.}
    \label{fig:fig13}
\end{center}
\end{figure}

 As for  $^{51}$Cr, we present in fig.~\ref{fig:fig12}  the
 comparison between the experimental level scheme and our calculations
 in the full space and with a $t=5$ truncation for the yrast band and
 the $\frac{1}{2}^-, \frac{3}{2}^-$ doublet around 2\,MeV. Again no
 relevant difference is appreciated between both calculated schemes
 and the agreement with the experiment is very good. Therefore, a
 $t=5$ truncation is used to get a more detailed spectroscopy
 for all the states below 3\,MeV and for the high spins up to
 $J=\frac{19}{2}^-$ (fig.~\ref{fig:fig13}). New data are available from
 a Gammasphere experiment aimed to the study of Coulomb energy
 differences in mirror nuclei~\cite{a51coul}. The agreement is very
 good in both regions, although the high spin states are slightly
 shifted up.

 Good agreement is also obtained for the electromagnetic moments of
 the ground state. The experimental values:

\begin{center}
$\mu_{exp}(\frac{5}{2}^-)=3.5683(13)\,\mu_N$\\
$Q_{exp}(\frac{5}{2}^-)=42(7)\,e\,fm^2$
\end{center}

\noindent
 are very well reproduced by the KB3G calculation:

\begin{center}
$\mu(\frac{5}{2}^-)=3.40\,\mu_N$\\
$Q(\frac{5}{2}^-)=35\,e\,fm^2$
\end{center}

  In addition, there is  experimental information on electromagnetic
  transitions, that we compare with the calculation in table~\ref{tab:mn51}.
  The accord is excellent for the M1 transitions. For the E2's the
  situation is similar to the other isobars; for the $\Delta J$=2
  transitions the agreement is quite good, while some discrepancies are
  present in the $\Delta J$=1 cases, that can be very large, for instance in
  the transition from $J=\frac{11}{2}^-$ to $J=\frac{9}{2}^-$.  As in
  the $^{51}$Cr case, we have found an inconsistency between the
  experimental values for the B(M1) and B(E2) and the experimental
  mixing parameter $\delta$. On the contrary the experimental value
  is fully compatible with the computed transition probabilities.
  Notice the abrupt change (more than two orders of magnitude
  decrease) in the transition probabilities  either M1 or E2
  of the $J=\frac{17}{2}^-$ state, that are spectacularly reproduced  by the
  calculation. A very intuitive physical
  explanation of this isomerism will be given later, when discussing
  the Coulomb energy
  differences between the yrast states of the mirror
   pair $^{51}$Fe\,-\,$^{51}$Mn.
\begin{table}[h]
\begin{center}
    \caption{Transitions in $^{51}$Mn.}
    \label{tab:mn51}
\vspace{0.3cm}
\begin{tabular*}{\textwidth}{@{\extracolsep{\fill}}ccc}
\hline\hline
& Exp. & Th.\\
\hline
$B(M1)$ & ($\mu_N^2$) & ($\mu_N^2$) \\
$\frac{7}{2}^-\rightarrow \frac{5}{2}^-$ & 0.207(34) & 0.177 \\
$\frac{9}{2}^-\rightarrow \frac{7}{2}^-$ & 0.16(5) & 0.114 \\
$\frac{11}{2}^-\rightarrow \frac{9}{2}^-$ & 0.6623(2148) & 0.423 \\
$\frac{17}{2}^-\rightarrow \frac{15}{2}^-$ & 0.00012(4) & 0.00003 \\
$\frac{19}{2}^-\rightarrow \frac{17}{2}^-$ & $>$0.5728 & 0.801 \\[0.8mm]
\hline
$B(E2)$ & ($e^2\,fm^4$) & ($e^2\,fm^4$) \\
$\frac{7}{2}^-\rightarrow \frac{5}{2}^-$ & 528(146) & 305 \\
$\frac{9}{2}^-\rightarrow \frac{5}{2}^-$ & 169(67) & 84 \\
$\frac{9}{2}^-\rightarrow \frac{7}{2}^-$ & 303(112) & 204 \\
$\frac{11}{2}^-\rightarrow \frac{7}{2}^-$ & 236(67) & 154 \\
$\frac{11}{2}^-\rightarrow \frac{9}{2}^-$ & 4.16(135) & 190 \\
$\frac{17}{2}^-\rightarrow \frac{13}{2}^-$ & 1.236(337) & 2.215\\[2mm]
\hline\hline
\end{tabular*}
\end{center}
\end{table}

\section{The isobars A=52}
\label{sec:espectro52}

 As in the $A=51$ isobaric multiplet we have performed full $0\hbar
 \omega$ calculations for all the isotopes in $^{52}$Cr and $^{52}$Mn 
 the full calculation is limited to  the yrast levels. More detailed
 spectroscopy is carried out with a $t$=5 truncation. The results of 
 the full $pf$-shell calculation for $^{52}$Fe have
 been already published in~\cite{fe52}. The experimental values are
 taken from ref.~\cite{nds}.

\subsection{Spectroscopy of $^{52}$Ca}
\label{sec:52ca}

\begin{figure}[h]
\begin{center}
    \epsfig{file=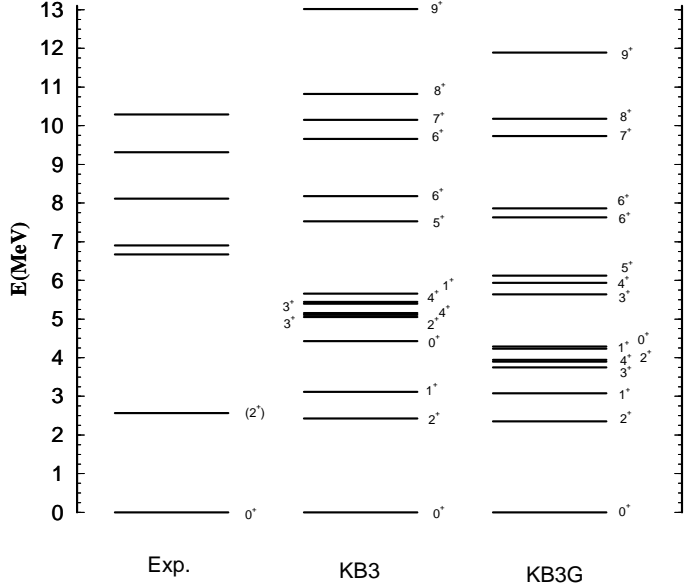,height=9cm}
    \caption{Energy levels of $^{52}$Ca.}
    \label{fig:fig14}
\end{center}
\end{figure}

  The experimental information is scarce, as can be seen in 
 figure~\ref{fig:fig14}. Both KB3 and KB3G
 give the same excitation energy for the doublet $J=1^+,2^+$ at 2.5~MeV,
 and reproduce correctly the energy of the first excited state $J=2^+$.
 The states whose leading configuration has four $p$-particles behave
 like the ground state under the KB3G changes, therefore we expect
 that their excitation energies do not change between KB3 and
 KB3G. This is the case for the above mentioned doublet, whose
 leading configuration is
 $f_{\frac{7}{2}}^8\,p_{\frac{3}{2}}^3\,p_{\frac{1}{2}}^1$ and for
 the second 
 0$^+$  whose configuration is
 $f_{\frac{7}{2}}^8\,p_{\frac{3}{2}}^2\,p_{\frac{1}{2}}^2$,
 therefore, both interactions
 locate them at the same place.

 The multiplet
 $J=1^+,2^+,3^+,4^+$ predicted at 5\,MeV by KB3 has a configuration
 $f_{\frac{7}{2}}^8\,p_{\frac{3}{2}}^3\,f_{\frac{5}{2}}^1$,
 consequently it is less affected by the monopole corrections 
 ($\Delta V_{pp}$) made in KB3G than the ground state, and  moves 
  down 1\,MeV in excitation energy.
 The states  $J=5^+$, $J=6^+_2$,  $J=7^+$,
 $J=8^+$ and $J=9^+$ are also lowered by KB3G, due to the occupation of
 the 1$f_{\frac{5}{2}}^1$ orbit.

\subsection{Spectroscopy of $^{52}$Sc}
\label{sec:52sc}

\begin{figure}[h]
\begin{center}
    \epsfig{file=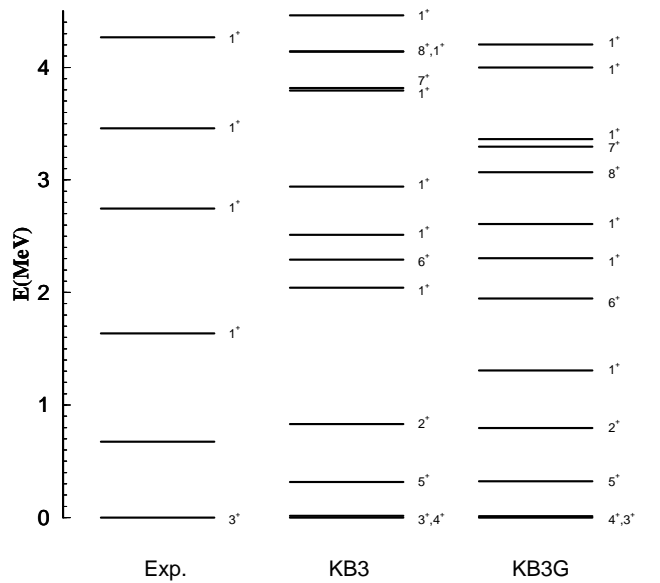,height=9cm}
    \caption{Energy levels of $^{52}$Sc.}
    \label{fig:fig15}
\end{center}
\end{figure}

 The dominant configuration in the ground state of $^{52}$Sc is
 $(f_{\frac{7}{2}}^1)_{\pi}(f_{\frac{7}{2}}^8\,p_{\frac{3}{2}}^3)_{\nu}$.
 It produces the multiplet of states with $J$ going from $2^+$ to
 $5^+$, below $1$\,MeV.  Experimentally little is known, the ground
 state 3$^+$, one state without spin assigned at about 0.6\,MeV and
 four 1$^+$ states seen in the beta decay of $^{52}$Ca~\cite{huck}
 (see fig.~\ref{fig:fig15}). In the figure we include the calculated
 ground state multiplet and the high spins up to $J=8^+$, as well as
 all the $J=1^+$ states present below 4.5\,MeV. According to the KB3G
 results, the state at 0.6\,MeV should be the 2$^+$ member of the
 ground state multiplet. In ref.~\cite{novo5} it is argued that this
 state could be a 1$^+$, based on the FPD6 results.  However this has
 to be attributed to the fact that FPD6 puts the 1$f_{\frac{5}{2}}$
 orbit too low at N=28.  If this state were a 1$^+$, it would have
 been seen in the decay of $^{52}$Ca, and this is not the case. The
 calculation produces six 1$^+$ states in the region where
 experimentally only four have been found. We will discuss the decay
 pattern of $^{52}$Ca in section~\ref{sec:decay52}.

\subsection{Spectroscopy of $^{52}$Ti}
\label{sec:52ti}

\begin{figure}[h]
\begin{center}
    \epsfig{file=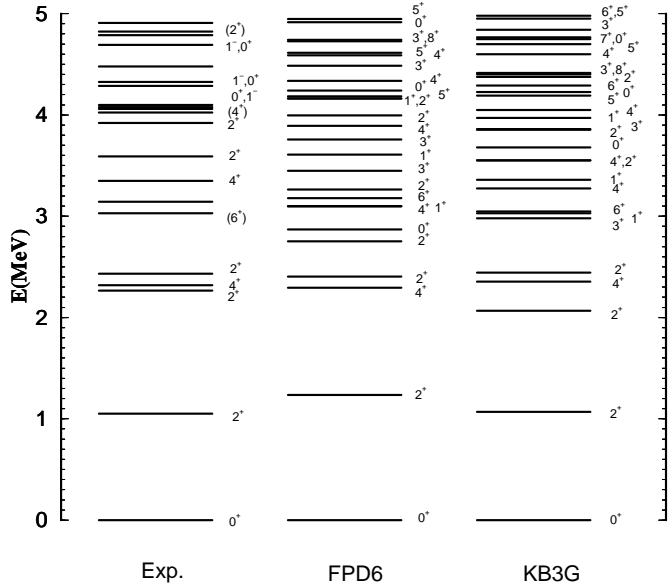,height=9cm}
    \caption{Energy levels of $^{52}$Ti.}
    \label{fig:fig16}
\end{center}
\end{figure}

The experimental level scheme is compared in figure~\ref{fig:fig16} with the
full pf-shell results using the interactions KB3G and FPD6.
The location of the yrast states from $J=0^+$ to $J=6^+$ corresponding to
the dominant configuration
$(f_{\frac{7}{2}}^8\,p_{\frac{3}{2}}^2)_\nu(f_{\frac{7}{2}}^2)_\pi$ is
correct for both interactions. The lowest states of $^{52}$Ti
can be described essentially in terms of the coupling $^{50}$Ti$\otimes
p_{\frac{3}{2}}^2$. Notice 
the right position of the triplet $J=2^+\,,4^+\,,2^+$ at 
$\sim 2.4$\,MeV. However, the doublet $J=4^+\,,2^+$ at $\sim 3.5$\,MeV is
clearly better placed by KB3G than by FPD6.
\begin{table}[h]
\begin{center}
\caption{Transitions in $^{52}$Ti.}
\label{tab:ti52}
\vspace{0.3cm}
\begin{tabular*}{\textwidth}{@{\extracolsep{\fill}}ccc}
\hline\hline
& Exp. & Th.\\
\hline
$B(M1)$ & ($\mu_N^2$) & ($\mu_N^2$) \\
$2_2^+\rightarrow 2_1^+$ & $0.55^{+0.41}_{-0.25}$ & 0.85 \\
$2_3^+\rightarrow 2_1^+$ & $>$0.16 & 0.51 \\[0.8mm]
\hline
$B(E2)$ & ($e^2\,fm^4$) & ($e^2\,fm^4$) \\
$2_1^+\rightarrow 0^+$ & $138^{+104}_{-92}$ & 85 \\
$2_2^+\rightarrow 0^+$ & $31^{+23}_{-14}$ & 16 \\
$2_3^+\rightarrow 2_1^+$ & $>$127 & 66 \\
$6^+\rightarrow 4^+$ & 123(22) & 80\\[2mm]
\hline\hline
\end{tabular*}
\end{center}
\end{table}
These states are dominantly
$f_{\frac{7}{2}}^8 f_{\frac{5}{2}}^2$ and they are shifted down by FPD6. This
is even more evident for the second $J=0^+$ state that is reasonably
positioned by KB3G but strongly shifted down by FPD6. The richness
of the calculated level scheme calls for an improvement of the
experimental spectrum.

There are six experimental transitions known~\cite{toi}, though
with large uncertainties. 
The calculated values are in accord with the data, as can be seen in
 table~\ref{tab:ti52}.

\subsection{Spectroscopy of $^{52}$V}
\label{sec:52v}

\begin{figure}[h]
\begin{center}
    \epsfig{file=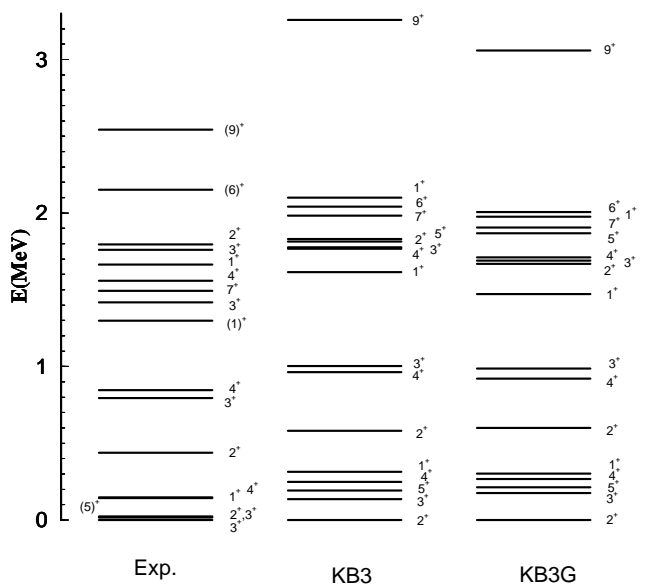,height=9cm}
    \caption{Energy levels of $^{52}$V.}
    \label{fig:fig17}
\end{center}
\end{figure}

 In figure~\ref{fig:fig17} we have plotted all the states with parity
 and spin assignments up to 2~MeV. Above this energy only the levels
 with spins equal or greater than $6^+$ are shown.  The situation is
 similar to $^{52}$Sc.  The dominant configuration
 $(f_{\frac{7}{2}}^3)_{\pi}(f_{\frac{7}{2}}^8\,p_{\frac{3}{2}}^1)_{\nu}$
 in the ground state produces the multiplet of states with $J$ going
 from $1^+$ to $5^+$. 

 The calculated ground state multiplet is more
 dilated that its experimental counterpart. Although the more
 relevant features are well accounted for we do not predict the
 correct ground state spin.  The $J=3^+,4^+$ doublet around $850$\,keV
 comes out inverted.  These states can be described as the result of 
 the coupling
 $^{51}$V$\otimes p_{\frac{3}{2}}^1$.  The high spin states $J=7^+$
 and $J=9^+$ are shifted up, making the agreement with experiment
 worse than in the  other nuclei studied in this work.

Only two electromagnetic transitions are experimentally
known~\cite{toi}: the E2  connecting the states $J=9^+$ and $J=7^+$
($B(E2)=81(6)\,e^2\,fm^4$) and the $M1$  conecting the first
excited  $J=2^+$ state 
and the  $J=3^+$  ground state ($B(M1)=1.7(4)\,\mu_N^2$). For the
first one the calculation produces a value  $B(E2)=87\,e^2\,fm^4$
and for the second $B(M1,2^+\rightarrow 3^+)=1.7\,\mu_N^2$,
a rather impresive agreement.

\subsection{Spectroscopy of $^{52}$Cr}
\label{sec:52cr}

\begin{figure}[h]
\begin{center}
    \epsfig{file=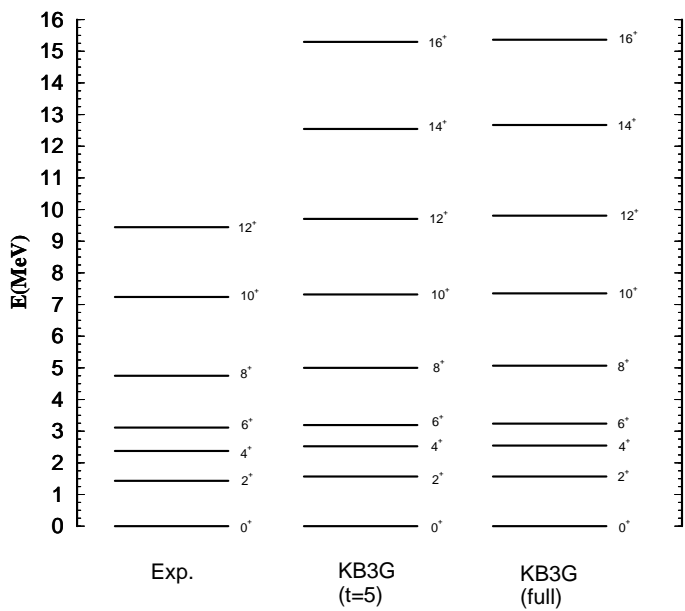,height=9cm}
    \caption{Yrast band of $^{52}$Cr; experiment, $t=5$ 
    and  full  calculation.}
    \label{fig:fig18}
\end{center}
\end{figure}

Figure~\ref{fig:fig18} shows the yrast states of $^{52}$Cr, up to
the band termination, calculated in the full $pf$-shell and in a $t=5$
truncation, compared with the experimental values. It is quite
evident that the states obtained with a $t=5$ truncation are nearly
converged. The accord of the KB3G results with the experiment is
excellent.

\begin{figure}[h]
\begin{center}
    \epsfig{file=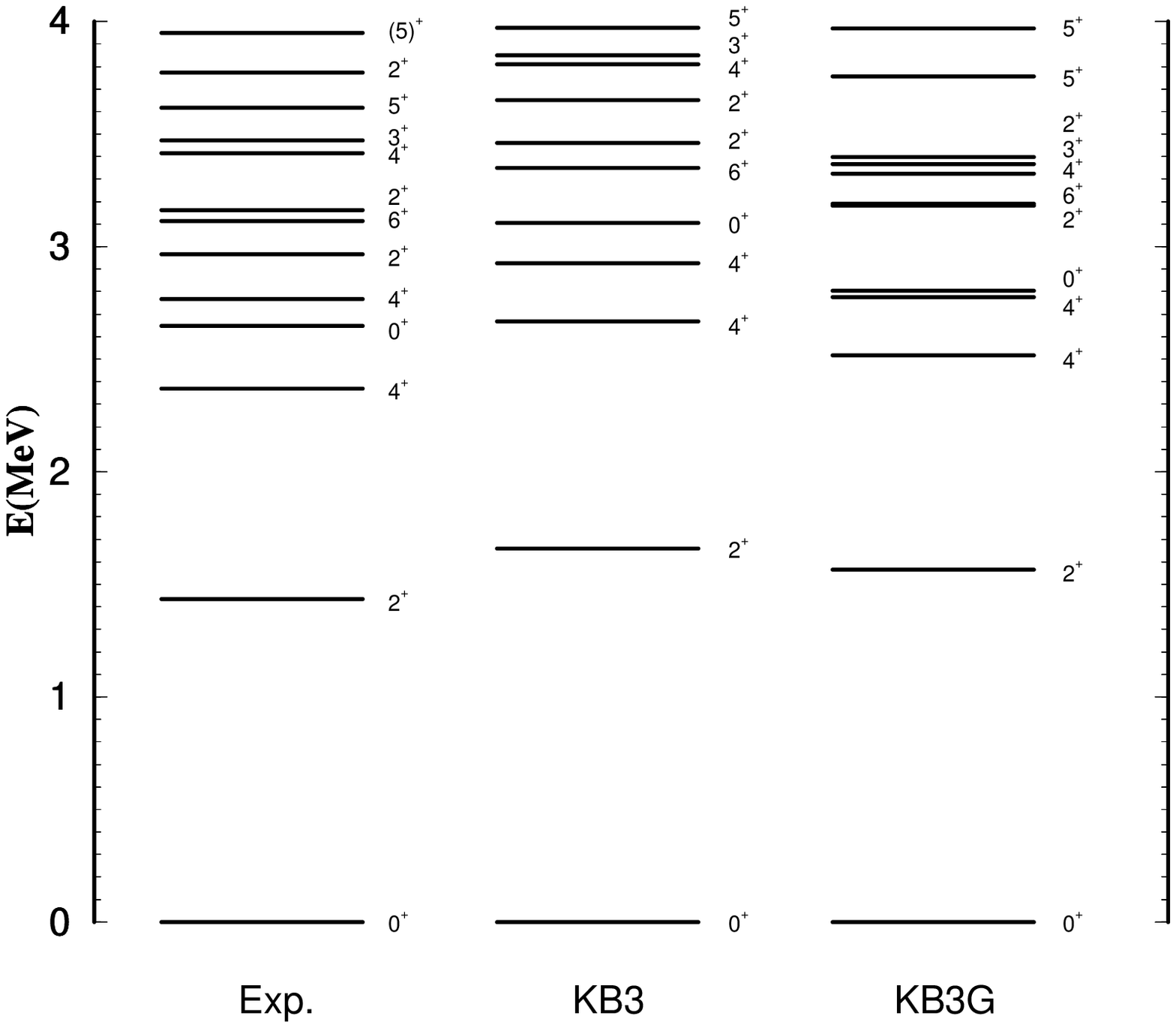,height=9cm}
    \caption{Energy levels of $^{52}$Cr up to 4\,MeV, $t=5$.}
    \label{fig:fig19}
\end{center}
\end{figure}

\begin{figure}[h]
\begin{center}
    \epsfig{file=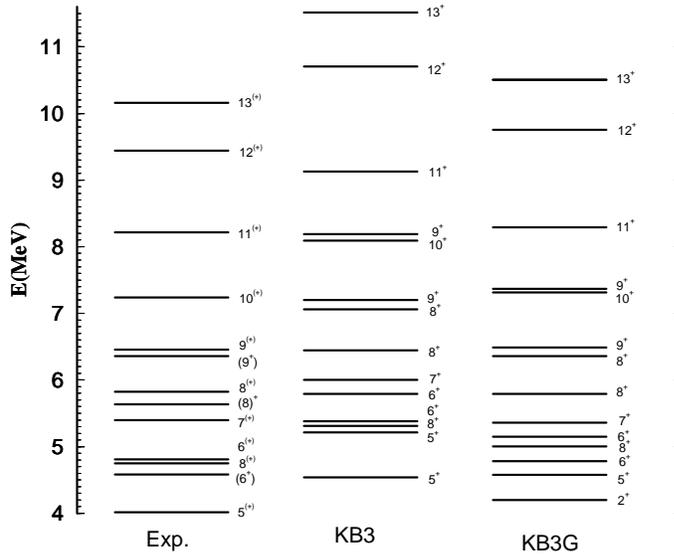,height=9cm}
    \caption{Energy levels of $^{52}$Cr above 4\,MeV, $t=5$.}
    \label{fig:fig20}
\end{center}
\end{figure}

 More detailed
spectroscopic results are shown in figures~\ref{fig:fig19}
and~\ref{fig:fig20} in a $t=5$ truncated calculation.  In
fig.~\ref{fig:fig19} all the states experimentally known up to
$\sim$~4\,MeV~\cite{nds} have been included as well as the
corresponding ones obtained with KB3G and KB3. The levels above this
energy are plotted in figure~\ref{fig:fig20}, where the experimental
data come from a recent experiment~\cite{appelbe}.  The agreement for
the non yrast states is also very satisfactory.  Notice the good
positioning of the doublets $J=0^+,4^+$ at 2.7\,MeV and $J=4^+,3^+$ at
3.5\,MeV and the one to one correspondence between the three $J=2^+$
and the three $J=5^+$ states measured in the excitation energy range
3\,MeV$\sim$~4\,MeV and the calculated ones.  
Beyond
4\,MeV the yrare states and the odd yrast are also fairly well
accounted for. 
It is clear from fig.~\ref{fig:fig20}
that KB3G produces a more satisfactory level scheme than KB3, that
shifts up  all the levels with spins greater than $J=8^+$, the maximum 
value attainable within the $(1f_{7/2})^{12}$ configuration. This is again a
manifestation of the too large N=28 gap produced by this interaction.    

The  experimental magnetic and quadrupole moments for the first
excited state $J=2^+$ are:

\begin{center}
$\mu_{exp}(2^+)=+3.00(50)\,\mu_N$\\
$Q_{exp}(2^+)=-8.2(16)\,e\,fm^2$
\end{center}

\noindent
and the calculated ones:

\begin{center}
$\mu(2^+)=+2.50\,\mu_N$\\
$Q(2^+)=-9.4\,e\,fm^2$
\end{center}

In table~\ref{tab:cr52} the known experimental values for the
electromagnetic transitions between the states of the yrast sequence are
given. The correspondence with the predicted ones is good with the
exception of the E2 transition $J=4^+$ to $J=2^+$, whose measured value
is far outside the range expected in this mass region.
Given its very large error bar, we would rather disregard this measure.
\begin{table}[h]
\begin{center}
\caption{Transitions in $^{52}$Cr.}
\label{tab:cr52}
\vspace{0.3cm}
\begin{tabular*}{\textwidth}{@{\extracolsep{\fill}}ccc}
\hline\hline
& Exp. & Th.\\
\hline
$B(M1)$ & ($\mu_N^2$) & ($\mu_N^2$) \\
$9^+\rightarrow 8^+$ & 0.05728(3759) & 0.040 \\[0.8mm]
\hline
$B(E2)$ & ($e^2\,fm^4$) & ($e^2\,fm^4$) \\
$2^+\rightarrow 0^+$ & 131(6) & 132 \\
$3^+\rightarrow 2^+$ & 7$^{+7}_{-5}$ & 5 \\
$4^+\rightarrow 2^+$ & 761(265) & 107 \\
$6^+\rightarrow 4^+$ & 59(2) & 68 \\
$8^+\rightarrow 6^+$ & 75(24) & 84 \\
$9^+\rightarrow 8^+$ & 0.5(20) & 0.6\\[2mm]
\hline\hline
\end{tabular*}
\end{center}
\end{table}

\subsection{Spectroscopy of $^{52}$Mn}
\label{sec:52mn}

\begin{figure}[h]
\begin{center}
    \epsfig{file=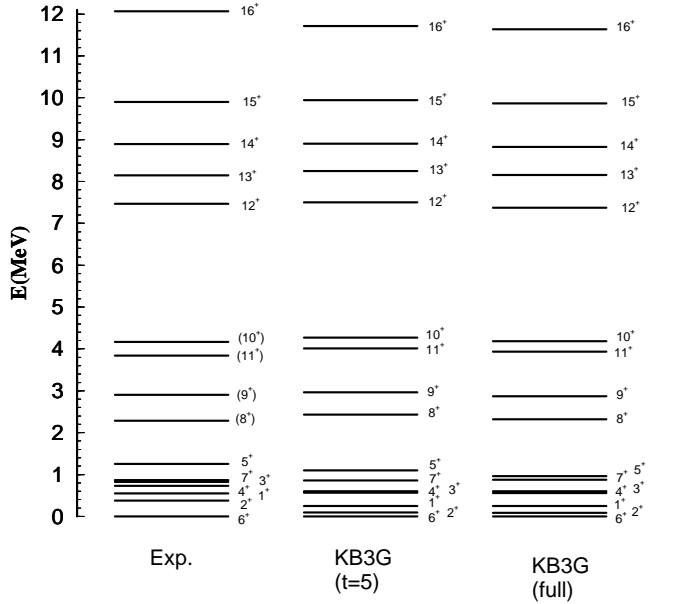,height=9cm}
    \caption{Yrast band of  $^{52}$Mn; experiment,  $t=5$ 
    and  full  calculation.}
    \label{fig:fig21}
\end{center}
\end{figure}

The full space calculation for the yrast states of the odd-odd nucleus
$^{52}$Mn produces only minor  differences  with the $t=5$
results (fig.~\ref{fig:fig21}). Notice the full correspondence
between the states of the multiplet below 1\,MeV. The experimental
data for the spins beyond the band termination ($11^+$ to $16^+$)
come from a recent experiment~\cite{silmn}. The agreement between
experiment and theory is spectacular. 
There is only three experimental states
with spin assignment not represented in the figure:
 a second $J=(5^+)$ at 1.42\,MeV, a
third $J=(5^+)$ at 1.68\,MeV and a second $J=(6^+)$ at 1.96\,MeV. The
 $t=5$ calculation places the second $J=5^+$ at 1.47\,MeV,
the third $J=5^+$  at 2.05\,MeV and the second $J=6^+$ at 1.91\,MeV.
 
The experimental magnetic and quadrupole  moments for the ground
state are known:

\begin{center}
$\mu_{exp}(6^+)=+3.0622(12)\,\mu_N$\\
$Q_{exp}(6^+)=+50(7)\,e\,fm^2$
\end{center}

the calculated values reproduce nicely these values:

\begin{center}
$\mu(6^+)=+2.9518\,\mu_N$\\
$Q(6^+)=+50\,e\,fm^2$
\end{center}

\noindent

Some electromagnetic transitions along the
yrast sequence have been measured. The calculations are in fair
agreement with them (see table~\ref{tab:mn52}).
\begin{table}[h]
\begin{center}
\caption{Transitions in $^{52}$Mn.}
\label{tab:mn52}
\vspace{0.3cm}
\begin{tabular*}{\textwidth}{@{\extracolsep{\fill}}ccc}
\hline\hline
& Exp. & Th.\\
\hline
$B(M1)$ & ($\mu_N^2$) & ($\mu_N^2$) \\
$7^+\rightarrow 6^+$ & 0.5012(2506) & 0.667 \\
$8^+\rightarrow 7^+$ & $>$0.015931 & 0.405 \\
$9^+\rightarrow 8^+$ & 1.074$^{+3.043}_{-0.537}$  & 0.759 \\[0.8mm]
\hline
$B(E2)$ & ($e^2\,fm^4$) & ($e^2\,fm^4$) \\
$7^+\rightarrow 6^+$ & 92$^{+484}_{-81}$ & 126 \\
$8^+\rightarrow 6^+$ & $>$1.15 & 33 \\
$8^+\rightarrow 7^+$ & $>$4.15 & 126 \\
$9^+\rightarrow 7^+$ & 104$^{+300}_{-46}$ & 66 \\
$11^+\rightarrow 9^+$ & 54(6) & 53\\[2mm]
\hline\hline
\end{tabular*}
\end{center}
\end{table}

\section{Half-lives and other $\beta$-decay properties}
\label{sec:decays}

Once the level schemes have been analysed, we study the $\beta ^-$
decays of these nuclei, to complete the  description of this mass
region. We present the results with the same ordering used for the
level schemes. Because of the increasing sizes of the calculations we
have limited this study to the isotopes of Calcium, Scandium, Titanium
and Vanadium. The calculated values for KB3 and KB3G gathered in the
tables correspond to  full
$0\hbar w$ calculation unless otherwise indicated.
 The $Q_{\beta^-}$ values are also included. The
errors attached to the calculated values proceed from  the errors in the
experimental $Q_{\beta^-}$ values. We compute the half-lives by making the
convolution of the strength function produced by the Lanczos method with
the Fermi function, increasing the number of iterations until
convergence is achieved.

\subsection{$\beta^-$ decays in the isobar chain $A=50$} 
\label{sec:decay50}

The experimental and calculated half-lives in the isobar chain $A=50$
are given in table~\ref{tab:vm50}. The agreement is quite
satisfactory. 
\begin{table}[h]
 \begin{center}
 \caption{Half-lives of  $A=50$ isobars. $Q_{\beta^-}$
values from~\protect\cite{nds}.}
 \label{tab:vm50}
\vspace{0.3cm}
 \begin{tabular*}{\textwidth}{@{\extracolsep{\fill}}cccccc}
\hline\hline
& & \multicolumn{3}{c}{$T_{\frac{1}{2}}$} &\\[0.2cm]
\cline{3-5}
A & $J^{\pi}$ & KB3 & KB3G & Exp. & $Q_{\beta^-}^{exp} (MeV)$\\ \hline
$^{50}Ca$ & $0^+$ & $11.1^{+0.3}_{-0.2}\,s$ & $11.2^{+}_{-}0.3\,s$ &
$13.9(6)\,s$ & $4.966(17)$\\

 & & & $12.3\pm0.3\,s$ & & \\
$^{50}Sc$ & $5^+$ & $130^{+}_{-}3\,s$ & $140^{+3}_{-1}\,s$ & $102.5(5)\,s$ &
 $6.888(16)$\\
 & & & $120^{+2}_{-1} \,s$ & & \\
$^{50}Ti$ & & \multicolumn{4}{c}{Stable} \\

$^{50}V$ & & \multicolumn{4}{c}{$4^{th}$ forbidden C.E.} \\[2mm]
\hline\hline
\end{tabular*}
\end{center}
\end{table}

\begin{figure}[h]
\begin{center}
    \epsfig{file=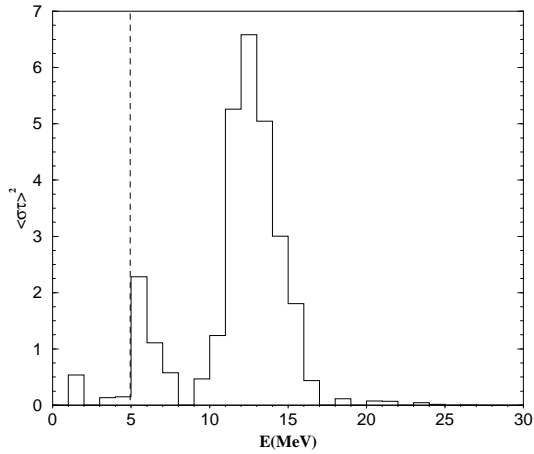,height=7cm}
    \caption{Gamow-Teller strength after 50 iterations for the
    $^{50}$Ca $\beta^-$ decay. The $<\sigma\tau>^2$ values are
    summed up in 1\,MeV bins. The dashed line indicates the
    experimental $Q_{\beta^-}$ value.}
    \label{fig:fig22}
\end{center}
\end{figure}

\begin{figure}[h]
\begin{center}
    \epsfig{file=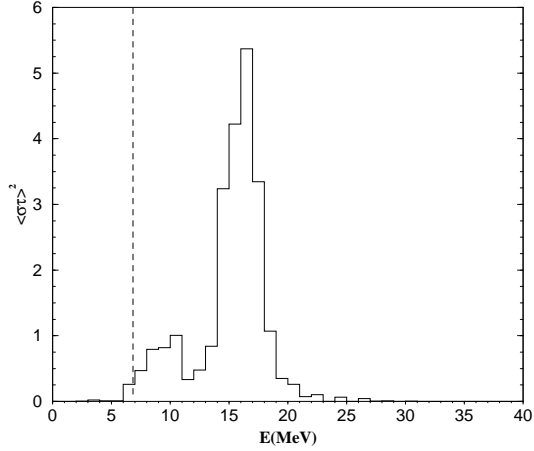,height=7cm}
    \caption{Gamow-Teller strength after 50 iterations for the
    $^{50}$Sc $\beta^-$ decay.  The $<\sigma\tau>^2$ values are
    summed up in 1\,MeV bins.  The dashed line indicates the
    experimental $Q_{\beta^-}$ value.}
    \label{fig:fig23}
\end{center}
\end{figure}

The Gamow-Teller strength function for $^{50}$Ca is shown in
 fig.~\ref{fig:fig22}. The amount of strength below the
$Q_{\beta^-}$ value is very small and it is concentrated at $\sim
 2$\,MeV. Notice that there is a large amount of
strength close above the $Q_{\beta^-}$ threshold. Thus, small
 variations in the excitation energies of the daughter states could
 influence (although not drastically) the final value of the half-life. 
In $^{50}$Sc
(see figure~\ref{fig:fig23}) the situation is similar.  Only
a tiny fraction of the total strength
lies below the $Q_{\beta^-}$ value.

\begin{figure}[h]
\begin{center}
    \epsfig{file=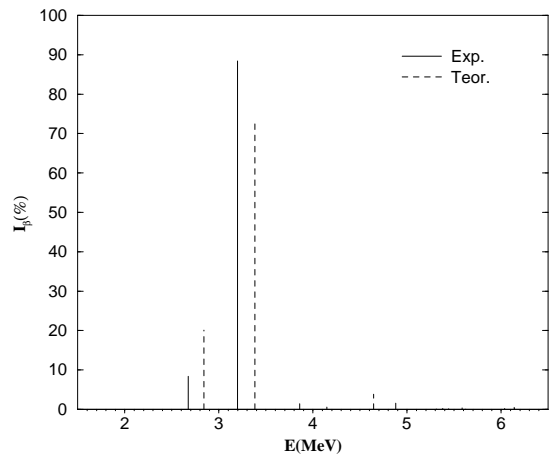,height=7cm}
    \caption{Percentages of the $\beta^-$ decay of
$^{50}$Sc. Experimental data from~\protect\cite{nds}.}
    \label{fig:fig24}
\end{center}
\end{figure}

We have computed the intensity (in percentage) of
the ground state decay to the  levels within the
$Q_{\beta^-}$ window of the daughter nucleus.  
$^{50}$Ca decays 99.7\% to the excited state $J=1^+$ at 1.78\,MeV in
$^{50}$Sc, what agrees perfectly with the experimental situation 
(99.0(13)\% decay to the $J=1^+$ state at 1.85\,MeV).  Analogously,
figure~\ref{fig:fig24} shows the experimental and calculated
intensities for the $\beta^-$ decay of $^{50}$Sc. Both experimental
peaks are reproduced by the calculation although they are slightly shifted.
Since the $^{50}$Ca decay feeds an unique state and $^{50}$Sc decays
to just two, it is possible to use the experimental energies of the
states to calculate the half-lives and to gauge the effect of the
phase space. Proceeding in such a way, the new calculated half-lives
for $^{50}$Ca and $^{50}$Sc are, respectively,
$T_{\frac{1}{2}}=12.3^{+}_{-}0.3\,s$ and
$T_{\frac{1}{2}}=120^{+2}_{-1}\,s$ which improve the agreement with
the experimental half-lives. These results are also shown  in
table~\ref{tab:vm50}.

\subsection{$\beta^-$ decays in the isobar chain $A=51$}
\label{sec:decay51}

Table~\ref{tab:vm51} gathers the experimental and calculated
half-lives for the isobar chain $A=51$. The calculated values are
quite close to the experimental ones.
\begin{table}[h]
\begin{center}
 \caption{Half-lives of $A=51$ isobars. $Q_{\beta^-}$
  values from~\protect\cite{nds}.}
 \label{tab:vm51}
\vspace{0.3cm}
\begin{tabular*}{\textwidth}{@{\extracolsep{\fill}}cccccc}
\hline\hline
& & \multicolumn{3}{c}{$T_{\frac{1}{2}}$} & \\[0.2cm]
\cline{3-5}
A & $J^{\pi}$ & KB3 & KB3G & Exp. &
$Q_{\beta^-}^{exp} (MeV)$\\ \hline

$^{51}Ca$ & $\frac{3}{2}^-$ & $8.9^{+1.0}_{-0.9}\,s$ &
$7.6^{+0.8}_{-0.7}\,s$ & $10.0(8)\,s$ &
$7.332(93)$\\

$^{51}Sc$ & $\frac{7}{2}^-$ & $14.1^{+}_{-}0.3\,s$ &  $10.4^{+}_{-}0.2\,s$
& $12.4(1)\,s$ & $6.508(20)$ \\

& & & $12.3\pm0.3\,s$ & & \\

$^{51}Ti$ & $\frac{3}{2}^-$ & $6.94^{+}_{-}0.02\,min$ &
$8.04^{+}_{-}0.02\,min$ & $5.76(1)\,min$ &
$2.4706(15)$ \\

& & & $6.97\pm0.02\,min$ & & \\

$^{51}V$ & & \multicolumn{3}{c}{Stable} &\\[2mm]
\hline\hline
\end{tabular*}
\end{center}
\end{table}

\begin{figure}[h]
\begin{center}
    \epsfig{file=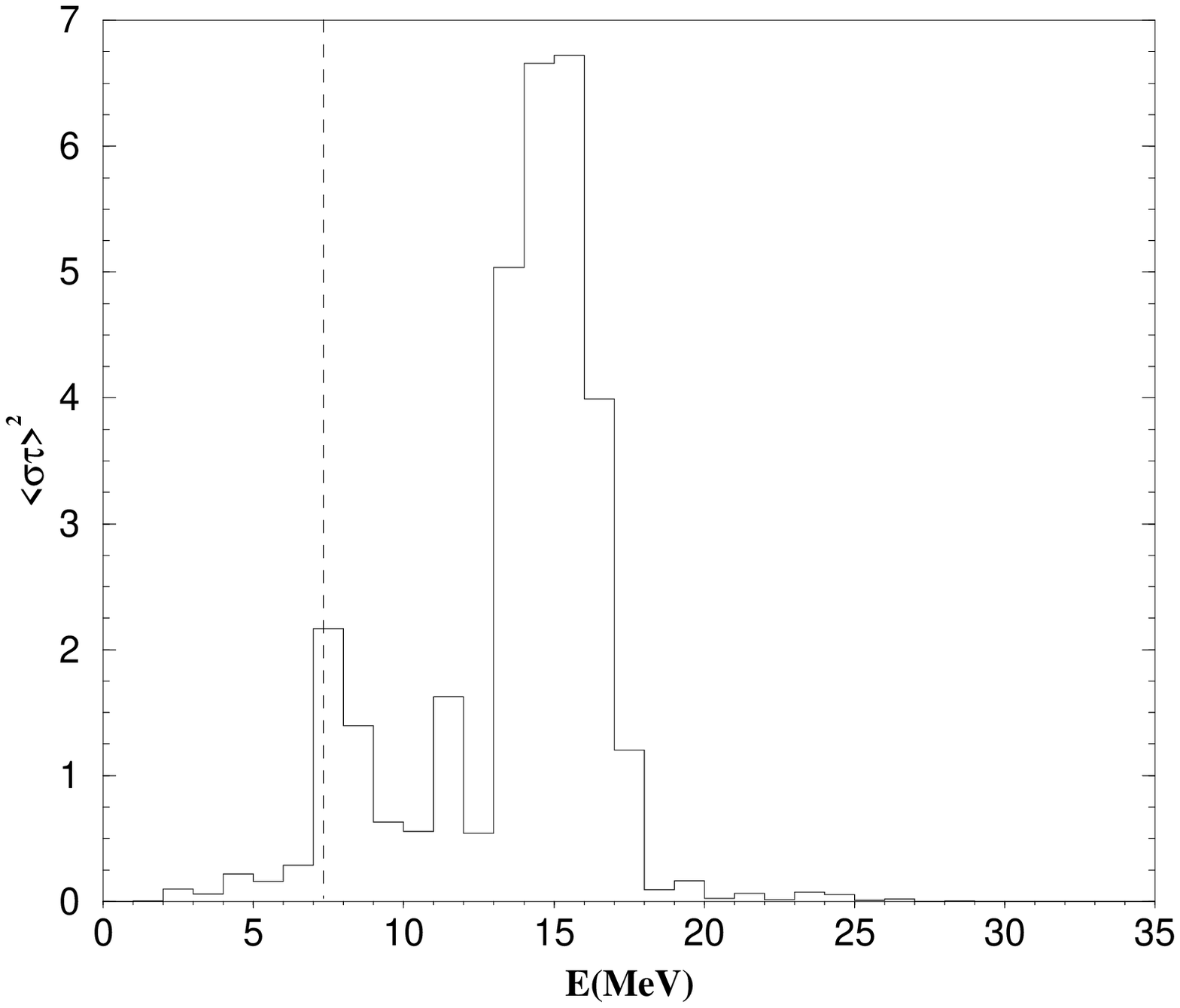,height=7cm}
    \caption{Gamow-Teller strength after 50 iterations for the
    $^{51}$Ca $\beta^-$ decay.  The $<\sigma\tau>^2$ values are
    summed up in 1\,MeV bins.  The dashed line indicates the
    experimental $Q_{\beta^-}$ value.}
    \label{fig:fig25}
\end{center}
\end{figure}

\begin{figure}[h]
\begin{center}
    \epsfig{file=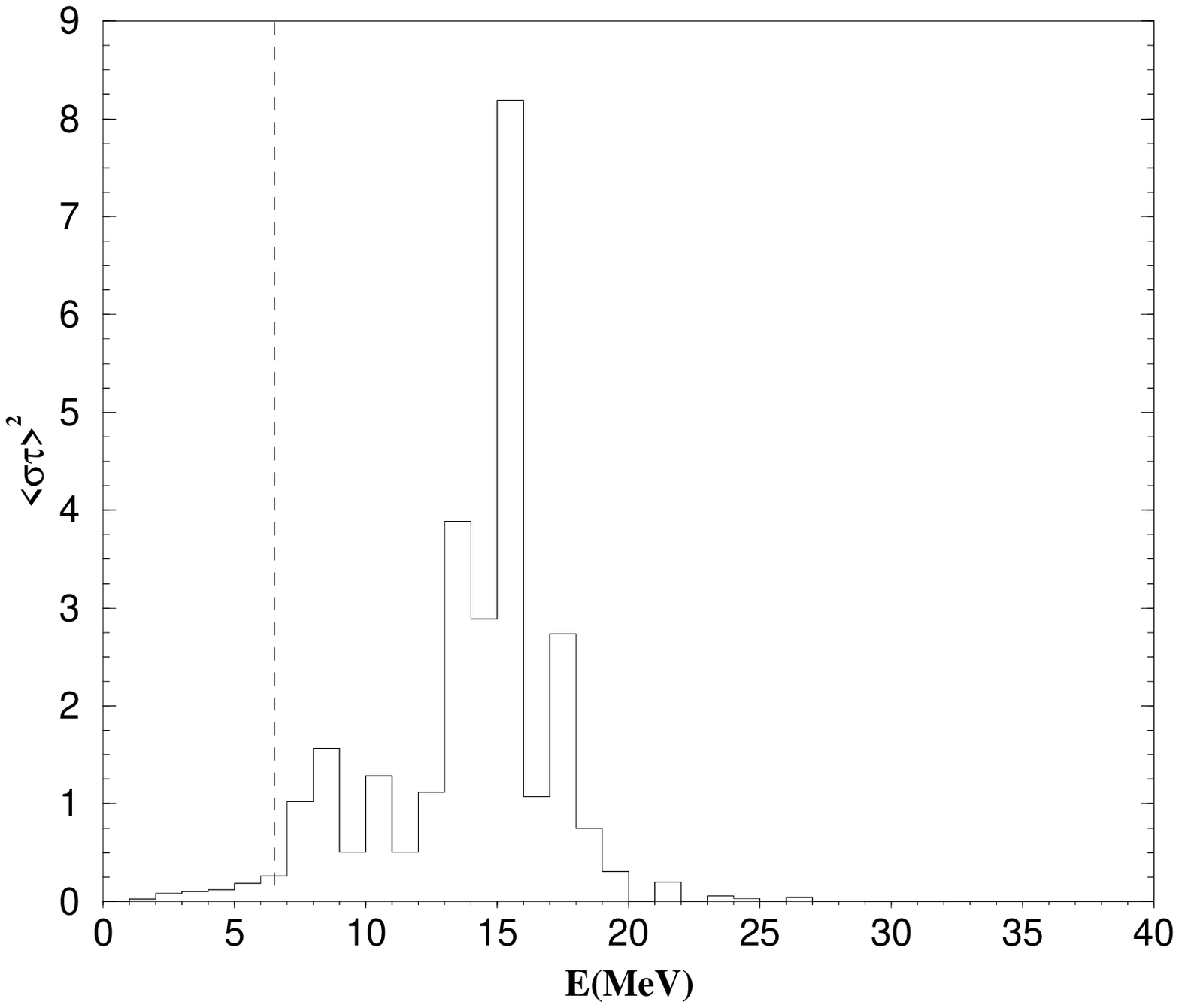,height=7cm}
    \caption{Gamow-Teller strength after 50 iterations for the
    $^{51}$Sc $\beta^-$ decay.  The $<\sigma\tau>^2$ values are
    summed up in 1\,MeV bins.  The dashed line indicates the
    experimental $Q_{\beta^-}$ value.}
    \label{fig:fig26}
\end{center}
\end{figure}

\begin{figure}[h]
\begin{center}
    \epsfig{file=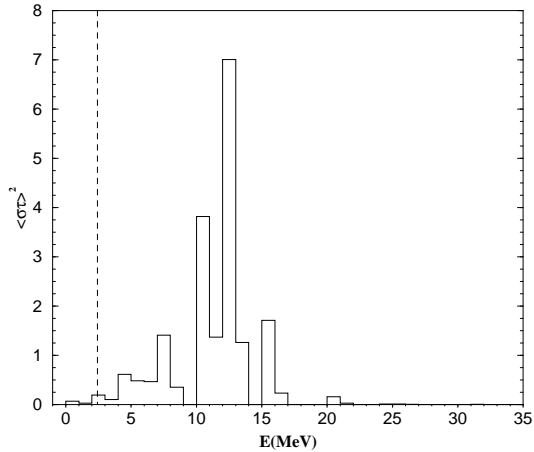,height=7cm}
    \caption{Gamow-Teller strength after 50 iterations for the
    $^{51}$Ti $\beta^-$ decay.  The $<\sigma\tau>^2$ values are
    summed up in 1\,MeV bins. The dashed line indicates the
    experimental $Q_{\beta^-}$ value.}
    \label{fig:fig27}
\end{center}
\end{figure}

Although within the $Q_{\beta^-}$ energy window there is a large
amount of Gamow-Teller strength (see figures~\ref{fig:fig25},~\ref{fig:fig26}
and~\ref{fig:fig27}), the situation looks like in the $A=50$
chain and much strenght is located around the $Q_{\beta^-}$ values.

\begin{figure}[h]
\begin{center}
    \epsfig{file=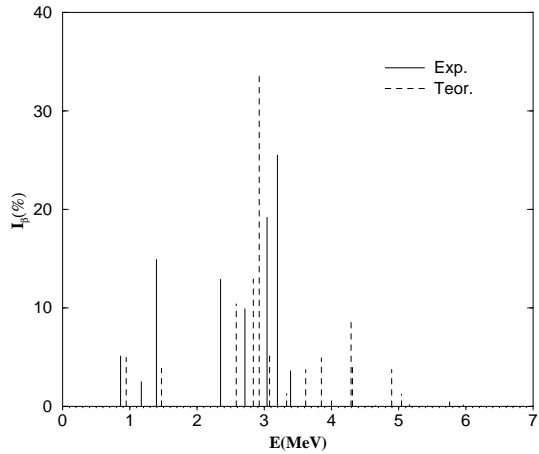,height=7cm}
    \caption{Percentages of the $\beta^-$ decay of
    $^{51}$Ca. Experimental data from~\protect\cite{nds}.}
    \label{fig:fig28}
\end{center}
\end{figure}

\begin{figure}[h]
\begin{center}
    \epsfig{file=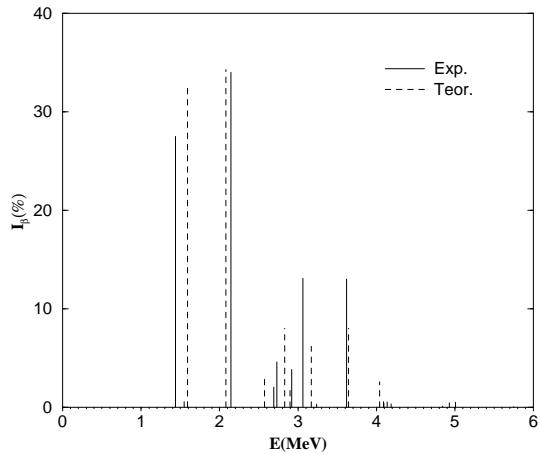,height=7cm}
    \caption{Percentages of the $\beta^-$ decay of
    $^{51}$Sc. Experimental data from~\protect\cite{nds}.}
    \label{fig:fig29}
\end{center}
\end{figure}

In figures~\ref{fig:fig28} and~\ref{fig:fig29} we compare the
experimental percentages for the decays with  the
calculation for both $^{51}$Ca and $^{51}$Sc. In the latter, the
agreement is very good.
$^{51}$Ca demands some comments. To compare with the experimental data
we must keep in mind that only states
up to 3.8\,MeV in $^{51}$Sc  have been observed.
In the first MeV  there is a single experimental state fed by
the decay. This fact is well reproduced by the calculation. 
Up to 3.5\,MeV there are seven  states experimentally
observed in the decay, the same number is given by the
calculation. However, the distribution of the intensity is not the
same. In the calculation all the intensity is concentrated between
2.5\,MeV and 3.0\,MeV, while experimentally there are two states strongly
fed at lower energy.

In the case of $^{51}$Ti the decay only feeds two states in the
daughter nucleus, a $J=\frac{3}{2}^-$ at 0.928\,MeV (8.1(4)\%) and a
$J=\frac{5}{2}^-$ at 0.320\,MeV (91.9(4)\%). The calculation agrees
with it since it predicts the decay to a $J=\frac{3}{2}^-$ state at
1.155\,MeV (6.5\%) and to another  $J=\frac{5}{2}^-$ state at
0.457\,MeV (93.5\%).

In view of the comparison of the intensities for the decays
 of $^{51}$Ca, $^{51}$Sc and
$^{51}$Ti, it seems plausible to make a new analysis in the same terms
we did in the previous section, i.e. to take the experimental
energies for the four states strongly fed in  $^{51}$Sc and the
corresponding two in  $^{51}$Ti. The new half-lives we
obtain  are $T_{\frac{1}{2}}=12.3^{+}_{-}0.3\,s$ for $^{51}$Sc and
$T_{\frac{1}{2}}=6.97^{+}_{-}0.02\,min$ for $^{51}$Ti, in much
better agreement with the experimental values shown in
table~\ref{tab:vm51}. For $^{51}$Ca this analysis is not so easy
since the number of states fed is larger and it
is not simple to make   a one to one correspondence between
 experimental and theoretical states.

\subsection{$\beta^-$ decays in the isobar chain $A=52$}
\label{sec:decay52}

In table~\ref{tab:vm52} the experimental data and the calculated
results for the half-lives of the nuclei in the $A=52$ isobar chain are shown.
For $^{52}$V we have only made a $t=4$ calculation, because, to be
consistent, the calculation in the daugther has to be $t=5$. Going
beyond that would have demanded an enormous computational effort.
The theoretical half-lives compare reasonably well with the experimental
numbers, although the discrepancy in $^{52}$Ca is severe. We shall now
analyse the intensity distributions and try to recompute the
half-lives using as much as possible the experimental excitation energies.
\begin{table}[h]
\begin{center}
 \caption{Half-lives of $A=52$ isobars.
  $Q_{\beta^-}$ values from~\protect\cite{nds}.}
 \label{tab:vm52}
\vspace{0.3cm}
\begin{tabular*}{\textwidth}{@{\extracolsep{\fill}}cccccc}
\hline\hline
& & \multicolumn{3}{c}{$T_{\frac{1}{2}}$} &\\[0.2cm]
\cline{3-5}
A & $J^{\pi}$ & KB3 & KB3G & Exp. & $Q_{\beta^-}^{exp} (MeV)$\\ \hline

$^{52}Ca$ & $0^+$ & $0.8^{+0.4}_{-0.3}\,s$ & $0.9^{+0.5}_{-0.3}\,s$ &
 $4.6(3)\,s$ & $7.900(500)$ \\

& & & $1.38^{+0.7}_{-0.4} \,s$ & & \\

$^{52}Sc$ & $3^+$ & $8^{+2}_{-1}\,s$ & $6.2^{+1.0}_{-0.8}\,s$ &
$8.2(2)\,s$ & $9.010(160)$\\

$^{52}Ti$ & $0^+$ & $2.69^{+}_{-}0.04\,min$ & $3.38^{+}_{-}0.07\,min$ &
$1.7(1)\,min$ & $1.973(8)$ \\

& & & $2.30\pm0.04\,min$ & & \\

$^{52}V$ & $3^+$ & $5.544^{+0.013}_{-0.014}\,min$ &
$5.79^{+}_{-}0.01\,min$ & $3.743(5)\,min$ &
$3.9756(12)$ \\
 & & & $4.87\pm0.01\,min$ & & \\[2mm]
\hline\hline
\end{tabular*}
\end{center}
\end{table}

\begin{figure}[h]
\begin{center}
    \epsfig{file=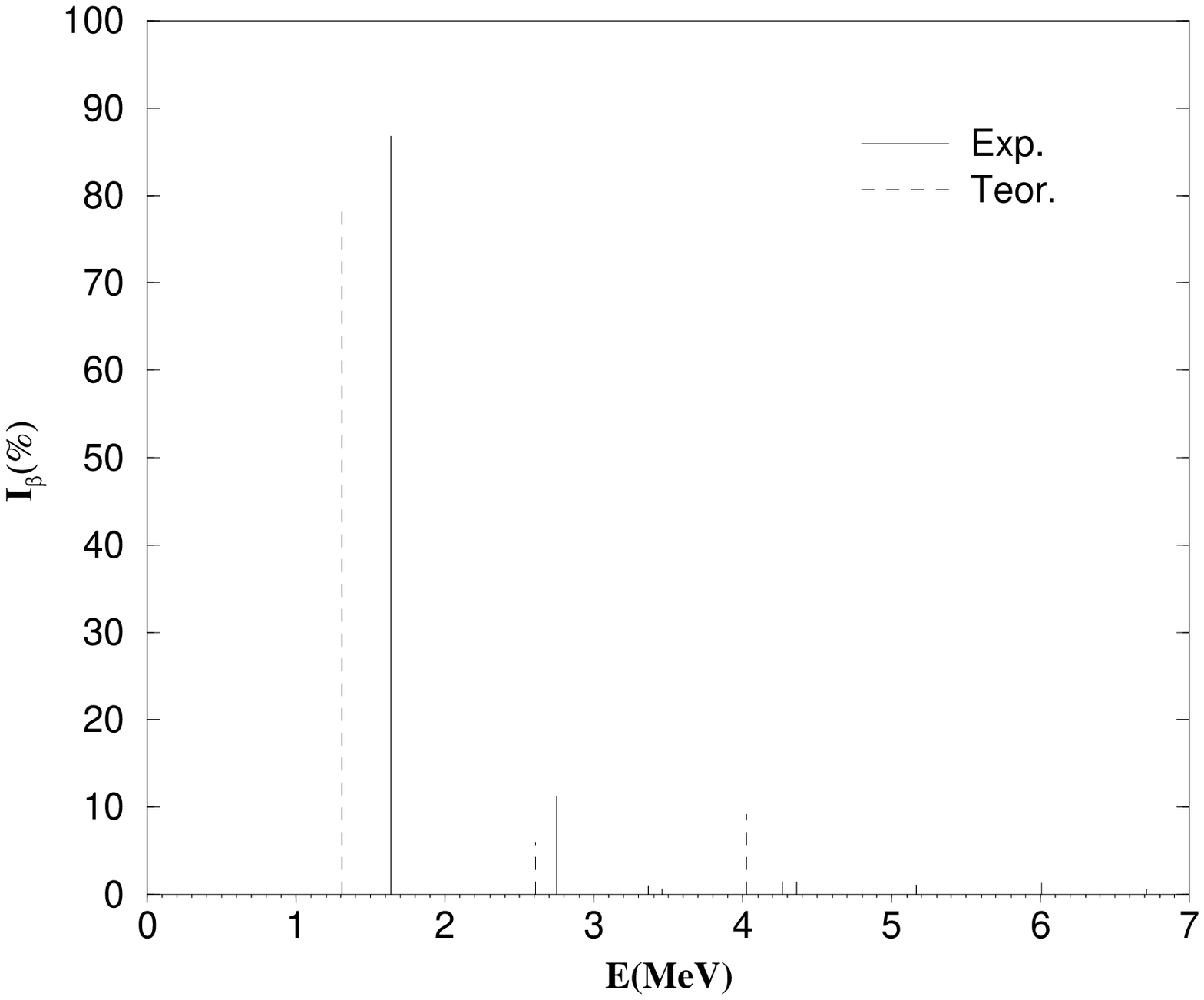,height=7cm}
    \caption{Percentages of the $\beta^-$ decay of
    $^{52}$Ca. Experimental data from~\protect\cite{nds}.}
    \label{fig:fig30}
\end{center}
\end{figure}

\begin{figure}[h]
\begin{center}
    \epsfig{file=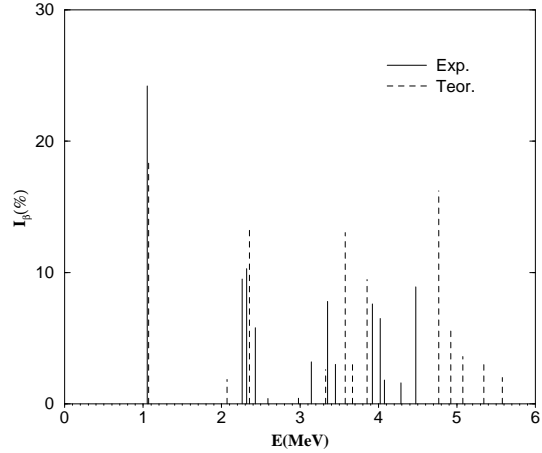,height=7cm}
    \caption{Percentages of the $\beta^-$ decay of
    $^{52}$Sc. Experimental data from~\protect\cite{nds}.}
    \label{fig:fig31}
\end{center}
\end{figure}

Figure~\ref{fig:fig30} shows the percentages for the $^{52}$Ca
$\beta^-$ decay and figure~\ref{fig:fig31} the same for 
$^{52}$Sc. The $^{52}$Ca case demands a special attention. When we
analyzed the $^{52}$Sc spectrum in section~\ref{sec:52sc} we mentioned
that there was one state without any experimental spin assignment at
675\,keV (fig.~\ref{fig:fig15}). In~\cite{novo5} the calculation was
made with the FPD6 interaction, producing a first $J=1^+$ at 701\,keV.
It was suggested that it could correspond to the 675\,keV state.  Our
interpretation of this state as the 2$^+$ member of the ground state
multiplet is experimentally supported by the fact that if this state
were a $J=1^+$, it would have been observed in the $\beta^-$ decay of
$^{52}$Ca. This is not the case, as can been seen in
figure~\ref{fig:fig30}. On the contrary, most of  the decay
(86.8\%) goes to the first experimental state with $J=1^+$
assignment, corresponding to the one we obtain in the calculation.
Thus, it is discarded that this unassigned experimental state could be
a $J=1^+$.
The first $J=1^+$ corresponds to a  
configuration with a neutron in the $f_{\frac{5}{2}}$ orbit. As we said in
the introduction, FPD6 locates the
$1f_{\frac{5}{2}}$ orbit too low, producing a very low
$J=1^+$ state.
The other three experimental $J=1^+$ states are also fed through
the decay of $^{52}$Ca although with smaller intensity. The
percentage of the second  at 2.7\,MeV is 11.2\% while for the third
at 3.5\,MeV  it is 0.6\% and 1.4\% for the last one at
4.3\,MeV. These data help us to make a one to one  correspondence 
between the states above 2\,MeV in figure~\ref{fig:fig15}. Our third
state $J=1^+$ at 2.6\,MeV corresponds to the second experimental one,
our fourth $J=1^+$ at 3.4\,MeV to the third  and the fourth
experimental $J=1^+$ to our sixth  at 4.4\,MeV.

The situation for $^{52}$Sc in figure~\ref{fig:fig31} is
very satisfactory. 
The decay to the $J=2^+$ excited state at 1\,MeV and to the triplet
around 2.3\,MeV is well reproduced, although in the calculation  the
triplet is more expanded than the experimental one. There is one
 state at 3.1\,MeV without assigned spin,  that seems to correspond to the
$3^+$ state near 3\,MeV (the calculation predicts a 0.7\% branch
to it). The next branch goes to the second $4^+$ at 3.3\,MeV.
Beyond, it becomes difficult to establish a detailed
correspondence.

The decay of  $^{52}$Ti feeds (100\%) a single state in the daughter
nucleus $^{52}$V, the $J=1^+$ at 141\,keV. This fact is reproduced
properly by the calculation which predicts a 100\% feeding of
the $J=1^+$ state at 300\,keV in the daughter nucleus for the
$\beta^-$ decay of $^{52}$Ti.
On the other hand, $^{52}$V decays  99.22(5)\%  to the $J=2^+$
excited state of $^{52}$Cr located at 1.4\,MeV, what agrees with the
calculation (a 98.92\% feeds the $J=2^+$ state at 1.5\,MeV).

Finally, we recalculate the half-lives using  the experimental
energies of the single state fed in the decay of  $^{52}$Ti and
$^{52}$V, and of the two states fed in the decay of  $^{52}$Ca. For
$^{52}$Sc the analysis in these terms is not so easy. The new
calculated half-lives are $T_{\frac{1}{2}}=2.30^{+}_{-}0.04\,min$ for
$^{52}$Ti and $T_{\frac{1}{2}}=4.87^{+}_{-}0.01\,min$ for
$^{52}$V. Both results improve the previous ones shown in
table~\ref{tab:vm52}.
For $^{52}$Ca  we get a half-life $T_{\frac{1}{2}}=1.38^{+0.7}_{-0.4}\,s$,
reducing substantially the initial discrepancy.

\section{Coulomb energy differences (CED) in the mirror pair
$^{51}$Mn-$^{51}$Fe}
\label{sec:ced}

It is well known that if the nucleon-nucleon interaction were charge symmetric
the  mirror nuclei  would have
the same level scheme. Thus, the small differences (normally a few tens of
keV) in their excitation energies, are due
to the isospin symmetry breaking Coulomb interaction.
The analysis of this effect has been recently pushed up in mass, with the
study of the  A=47 and A=49 mirror pairs \cite{bent47,bent49}. It
was found that the CED's are extremely sensitive to the structure of
the nuclear wave functions. The next step has been to measure 
the $^{51}$Mn-\,$^{51}$Fe mirror pair, the
heaviest one in which high spin states up to the band termination have
been observed~\cite{a51coul}.
\begin{figure}[h]
\begin{center}
    \epsfig{file=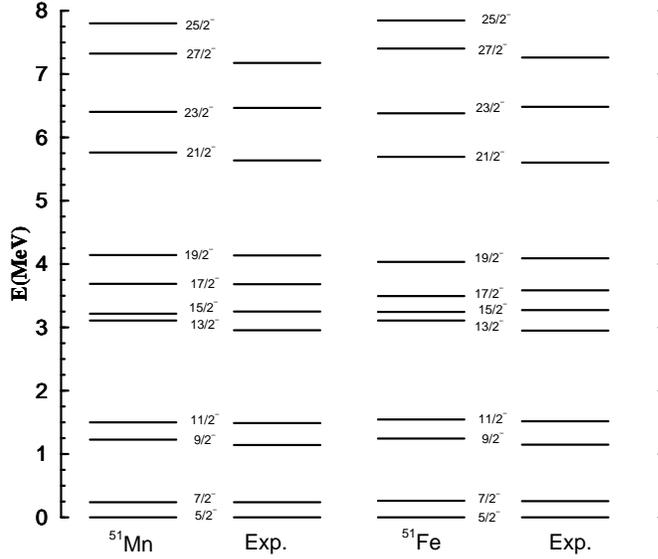,height=9cm}
    \caption{Excitation energies of  yrast  bands of the mirror pair
    $^{51}$Mn\,-\,$^{51}$Fe. KB3G in the full $pf$-shell.
     Coulomb included.}
    \label{fig:fig32}
\end{center}
\end{figure}

\begin{figure}[h]
\begin{center}
    \epsfig{file=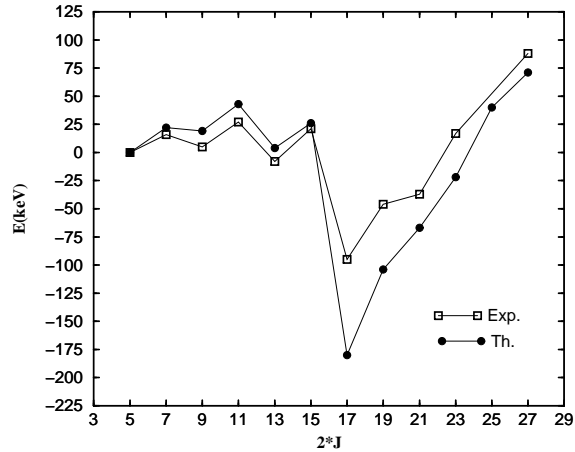,height=7cm}
    \caption{Experimental and calculated CED defined as
    $E_x(^{51}$Fe$)-E_x(^{51}$Mn$)$.}
    \label{fig:fig33}
\end{center}
\end{figure}

In figure~\ref{fig:fig32} the calculated spectra for both nuclei,
including the Coulomb interaction are  compared with the experimental
data. The Coulomb matrix elements are those denoted ``A42'' in
ref~\cite{bent49}.
At this scale, very slight differences (hardly any) are appreciable
between them.
However, looking more carefully at the CED's 
an abrupt change is observed
at $J=\frac{17}{2}^-$ (fig.~\ref{fig:fig33}). The effect is also present
in the calculation, that shows exactly the same trends as the
experiment, although with enhanced  values. The large increase in the
CED can be
interpreted as due to the alignment of one proton pair in $^{51}$Fe which
does not occur in $^{51}$Mn. As a consequence, the Coulomb energy is
sharply reduced in $^{51}$Fe but not in $^{51}$Mn, therefore 
the CDE increases dramatically  (in absolute value). The breaking of a
proton pair occurs in  $^{51}$Fe because it has an even number of
protons while the odd proton  in $^{51}$Mn blocks the breaking
process. In $^{51}$Mn, a neutron pair is broken, but this  has
no effect in the Coulomb energy. Beyond $J=\frac{17}{2}^-$,
the protons start aligning  also in $^{51}$Mn and therefore
the CED's approach  zero at the band termination.

\begin{figure}[h]
\begin{center}
    \epsfig{file=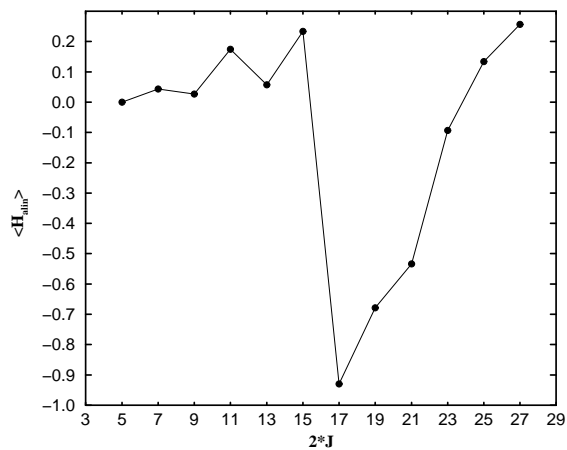,height=7cm}
    \caption{Contribution from $1f_{\frac{7}{2}}$ proton pairs coupled
     to the maximum spin $J=6$ (see text). $^{51}$Mn minus $^{51}$Fe.}
    \label{fig:fig34}
\end{center}
\end{figure}

In order to make more visible the proton pair breaking, we have
calculated the expectation value of the operator H$_{alin}=\left[ 
(a^+_{\frac{7}{2}}a^+_{\frac{7}{2}})^{6,1}_{\pi\pi}
(a_{\frac{7}{2}}a_{\frac{7}{2}})^{6,1}_{\pi\pi}\right]^0$
for the states of each nucleus. This operator acts as a sort of counter
for the  pairs of $1f_{\frac{7}{2}}$ protons coupled to the maximum spin value
$J=6$. In figure~\ref{fig:fig34} we have plotted the 
difference between its expectation values for the states of $^{51}$Mn
and $^{51}$Fe.
It is manifest  that  a proton pair fully
aligns  at $J=\frac{17}{2}^-$ in $^{51}$Fe and not at all
in $^{51}$Mn. 

The sudden change of structure at
$J=\frac{17}{2}^-$  should also show up in the 
electromagnetic transitions. Indeed, this is the case and the
general trend of the 
transition values up to $J=\frac{15}{2}^-$ breaks at
$J=\frac{17}{2}^-$ dropping to  nearly  zero. For the $E2$
transitions we have $B(E2,\frac{17}{2}^-\rightarrow \frac{15}{2}^-)
=0.013\,e^2\,fm^4$, $B(E2,\frac{17}{2}^-\rightarrow \frac{13}{2}^-)
=2.012\,e^2\,fm^4$ and for the $M1$ transition
$B(M1,\frac{17}{2}^-\rightarrow \frac{15}{2}^-) =0.00002\,\mu^2_N$, a
reduction of several orders of magnitude that makes the state
$\frac{17}{2}^-$ isomeric.

\section{Magnetic moments of the N=28 isotones}
\label{sec:mu}

New measures of the magnetic moments of the  N=28 isotones~\cite{speidel,ernst}
have brought up again the question of the sensitivity of these
observables to the effective interactions and to the valence space 
truncations as
well as the relevance of the use of bare or effective $g$-factors. We
have examined this issue using the new interaction KB3G and we compare
its results with those obtained using FPD6.  

The ground  and  low-lying states of the N=28 isotones are
dominated by  1$f_{7/2}^n$ configurations. In the even-even cases, what
is measured is the magnetic moment of the first excited 2$^+$,
therefore the closed shell nuclei  $^{48}$Ca and  $^{56}$Ni are excluded
from this systematics. In the $1f_{7/2}^n$ limit, the value of the
2$^+$ magnetic moment, which is the same for  $^{50}$Ti, $^{52}$Cr and
 $^{54}$Fe, is 3.31~$\mu_N$, using bare $g$-factors. The usual choice
 of effective $g$-factors, $g^s_{eff}$=0.75 $g^s_{bare}$,
  $g^l_{\pi}$=1.1~$\mu_N$ and  $g^l_{\nu}$=-0.1~$\mu_N$, gives
  3.08~$\mu_N$.
 For the odd isotones, the ground states have J=7/2 and  $\mu$=5.79~$\mu_N$
with bare $g$-factors and  $\mu$=5.39~$\mu_N$ with the effective
$g$-factors. The differences amount to 7\%. When mixing is allowed,
the tendency is to reduce the difference between the results using
bare and effective $g$-factors.

In table~\ref{tab:n28mu1} we have collected the results for the even
nuclei calculated with KB3G and FPD6 at different truncation
levels. For KB3G, a large reduction (20\%) occurs between t=0 and t=3,
 while a marginal 5\% more is obtained in the full calculation. For
 FPD6 the reductions are larger (30\% and 10\% respectively).
 The FPD6 results are too small
 compared with the experiment even if bare $g$-factors are used, while
 those of KB3G are reasonably close to the measured values. The reason
 for the too large reduction that FPD6 produces can be related to the
 larger mixing with the  1f$_{5/2}$ orbit, due to its very low
 location that we have already discussed.
 The experimental values of $\mu$ show a slight tendency to decrease
 with the number of protons, a trend that is weaker or even absent in
 the theoretical predictions.

\begin{table}[h]
\begin{center}
\caption{Magnetic moments of the  first excited 2$^+$ states of the 
 even N=28 isotones (in $\mu_N$).}
\label{tab:n28mu1}
\vspace{0.3cm}
\begin{tabular*}{\textwidth}{@{\extracolsep{\fill}}cccc}
\hline\hline
     &$^{50}$Ti  & $^{52}$Cr   &$^{54}$Fe  \\
\hline
EXP  & 2.89(15)   & 2.41(13)    & 2.10(12)  \\
\hline
KB3G(full)(bare) & 2.52   & 2.50    &        \\
KB3G(t=3)(bare) & 2.65   & 2.67    & 2.56     \\
KB3G(t=3)(eff) & 2.49   & 2.57    & 2.52     \\
\hline
FPD6(full)(bare) & 2.28 & 1.90 &    \\
FPD6(full)(eff) & 2.12 & 1.87 &    \\
FPD6(t=5)(bare) & 2.30 & 2.08 & 2.10  \\
FPD6(t=5)(eff) & 2.14 & 2.04 & 2.12  \\
FPD6(t=3)(bare) & 2.49 & 2.38 & 2.26  \\
FPD6(t=3)(eff) & 2.32 & 2.32 & 2.29  \\
\hline\hline
\end{tabular*}
\end{center}
\end{table}

  In table~\ref{tab:n28mu2} we present the results for the odd
  isotones. In this case KB3G seems to reproduce the decreasing trend
  better than FPD6, that gives a flatter behaviour. Considering the
  expected reductions in going from t=3 to the full calculation, the
  agreement can be considered quite decent, although in this case the
  use of effective $g$-factors worsens it. As in the case of the even
  isotones, FPD6 gives too small values.   

\begin{table}[h]
\begin{center}
\caption{Magnetic moments of the 7/2$^-$ ground  states of the 
 odd  N=28 isobars (in $\mu_N$).}
\label{tab:n28mu2}
\vspace{0.3cm}
\begin{tabular*}{\textwidth}{@{\extracolsep{\fill}}ccccc}
\hline\hline
     &$^{49}$Sc  & $^{51}$V   &$^{53}$Mn & $^{55}$Co  \\
\hline
EXP  &   --   & 5.15    & 5.02(1) &  4.822(3) \\
\hline
KB3G(full)(bare) & 5.34    &  4.99    &   &   \\
KB3G(t=3)(bare) & 5.30  & 5.05    & 4.91   & 4.75 \\
KB3G(t=3)(eff) & 5.00  & 4.80  & 4.71  & 4.61  \\
\hline
FPD6(t=3)(bare) & 5.09 & 4.89 & 4.88 & 4.81 \\
FPD6(t=3)(eff) & 4.77 & 4.65 & 4.67 & 4.63 \\
\hline\hline
\end{tabular*}
\end{center}
\end{table}

\section{Conclusions}
 In this work we have  extended our previous full
 $pf$-shell studies of the A=48 and A=49 isobars three units of
 mass. In order to be able to treat properly the N=Z=28 shell closure
 and its surroundings we have introduced a mass dependence in the 
 interaction KB3 and refined its original monopole changes. This results in
 the KB3G interaction. G emphasizes that the new interaction produces the
 right quasiparticle gaps for all the N=28 isotones. With this interaction
 --that can be interpreted as a natural extension of KB3, equivalent
 to it for the nuclei studied already-- we achieve an extremely high
 quality description of  the abundant
 experimental data available in the mass region A=50/51/52.
 Besides, we are now in a much better situation
 concerning our predictive power for higher masses.
 We have also studied the beta decay of the isotopes
 of Ca, Sc, V and Ti, computing half-lives and strength distributions,
 that agree nicely with the experimental information too.
 In the final sections we dwell on the magnetic
 moments of the N=28 isotones and on the  Coulomb displacement
 energies along the yrast band of
 the A=51 mirror nuclei.

\ack
 This work has been partly supported by a grant of the 
 DGES~(Spain), ref. PB96-053 and by the IN2P3~(France)-CICyT~(Spain)
 agreements. We also thank the CCCFC-UAM for a computational grant.

\end{document}